\documentclass[prb,twocolumn,showpacs]{revtex4}
\usepackage{graphicx}
\usepackage{amsmath}

\begin{document}
\title{Selection rules for Single-Chain-Magnet behavior in non-collinear Ising
systems}
\author{Alessandro Vindigni$^1$}
\email{vindigni@phys.ethz.ch}
\author{Maria Gloria Pini$^2$} \affiliation{$^1$Laboratorium f\"ur
Festk\"orperphysik, Eidgen\"ossische Technische Hochschule
Z\"urich, CH-8093 Z\"urich, Switzerland
\\
$^2$Istituto dei Sistemi Complessi, Consiglio Nazionale delle
Ricerche, Via Madonna del Piano 10, I-50019 Sesto Fiorentino (FI),
Italy}
\date{\today}
\begin{abstract}
The magnetic behavior of molecular Single-Chain Magnets is investigated 
in the framework of a one-dimensional
Ising model with single spin-flip Glauber dynamics. Opportune
modifications to the original theory are required in order to
account for reciprocal non-collinearity of local anisotropy axes
and the crystallographic (laboratory) frame. The
extension of Glauber's theory to the case of a collinear Ising
ferrimagnetic chain is also discussed. Within this formalism, both
the dynamics of magnetization reversal in zero field and the
response of the system to a weak magnetic field, oscillating in
time, are studied. Depending on the geometry, selection rules
are found for the occurrence of slow relaxation of the
magnetization at low temperatures, as well as for resonant
behavior of the {\it a.c.} susceptibility as a function of
temperature at low frequencies. The present theory  applies successfully to some real
systems, namely Mn-, Dy-, and Co-based molecular magnetic chains,
showing that Single-Chain-Magnet behavior is not only a feature of
collinear ferro- and ferrimagnetic, but also of canted
antiferromagnetic chains.
\end{abstract}
\maketitle

\section{Introduction}
Slow dynamics of the magnetization reversal is a crucial
requirement for potential applications of Single-Chain Magnets
(SCM's)~\cite{Review_CCM,Miyasaka,feature_SCM}, and nanowires in
general, in magnetic-memory manufacture. For nanowires with a
biaxial anisotropy, provided that their length is much greater
than the cross section diameter but smaller than exchange length,
this phenomenon is governed by thermal nucleation and propagation
of soliton-antisoliton pairs; the associated characteristic time
is expected to follow an Arrhenius law~\cite{Braun1,Braun2}. For
genuine one dimensional (1D) Ising systems with single spin-flip
stochastic dynamics, a slow relaxation of the magnetization
was first predicted by Glauber~\cite{Glauber} in 1963.
Through Glauber's approach, many physical systems were
investigated, ranging from
dielectrics~\cite{Anderson,Bozemir,Bozemir1} to
polymers~\cite{Anderson,Skinner,Berim_Ruckenstein}. More
fundamentally, this model has been employed to justify the use of
the Kohlrauch-Williams-Watts function~\cite{Skinner,Brey_Prados}
(stretched exponential) to fit the relaxation of generalized 1D
spin systems. Also the universality issue of the dynamic critical
exponent~\cite{Stinchcombe,da_Silva2,da_Silva,Toboshnik,PiniRettori} of the
Ising model~\cite{Ising}, as well as strongly out-of-equilibrium
processes (magnetization reversal~\cite{Einax_Schulz}, facilitated
dynamics~\cite{Einax_Schulz}, etc.) have been studied moving
from the basic ideas proposed by Glauber.

In this paper, single spin-flip Glauber dynamics is used to
investigate theoretically the slow dynamics of the magnetization
reversal in molecular magnetic systems. In particular, we extend
Glauber's theory~\cite{Glauber} to the Ising collinear
ferrimagnetic chain, as well as to the case of a chain in which
reciprocal non-collinearity of local anisotropy axes and the
crystallographic (laboratory) frame is encountered. Such
extensions are motivated by the fact that (i) in molecular-based
realizations of SCM's, antiferromagnetic coupling typically has a
larger intensity than the ferromagnetic one; in fact, the
overlapping of magnetic orbitals, which implies antiferromagnetic
exchange interaction between neighboring spins, can be more easily
obtained than the orthogonality condition, leading to
ferromagnetism~\cite{ChemBond,Goodenough,Kanamori}; (ii)
non-collinearity between local anisotropy axes and the
crystallographic (laboratory) frame takes place quite often in
molecular spin chains.
Besides magnetization reversal,
the dynamic response of the system to a weak magnetic
field, oscillating in time at frequency $\omega$, is also
studied.  Depending on the specific experimental geometry, selection rules are
found for the occurrence of resonant behavior of the {\it a.c.}
susceptibility as a function of temperature (stochastic resonance)
at low frequencies, as well as for slow relaxation of the
magnetization in zero field at low temperatures.

The paper is organized as follows. In Sect. II we extend Glauber's
theory~\cite{Glauber}, originally formulated for a chain of
identical and collinear spins, to the more general model of a
chain with non-collinear spins, possibly with Land\'e factors that
vary from site to site. In Sect. III we use two different
theoretical methods (the Generating Functions approach and
the Fourier Transform approach) to investigate the relaxation of
the magnetization after removal of an external static magnetic
field, starting from two different initial conditions: fully saturated
or partially saturated. 
In Sect. IV we calculate, in a linear approximation, the magnetic
response of the system to an oscillating magnetic field.
For a chain of $N$ spins, the {\it a.c.}
susceptibility is expressed as the superposition of $N$
contributions, each characterized by its time scale; through a few
simple examples, we show that, depending on the geometry of the
system (\textit{i.e.}, the relative orientations of the local easy
anisotropy axes and of the applied field), different time scales
can be selected, possibly giving rise, for low frequencies, to a
resonant peak in the temperature-dependence of the complex magnetic
susceptibility.  In Sect. V we show that the theory provides a
satisfactory account for the SCM behavior
experimentally observed in some magnetic molecular chain
compounds, characterized by dominant antiferromagnetic exchange
interactions and non-collinearity between spins. Finally, in Sect. VI, the
conclusions are drawn and possible forthcoming applications are also discussed.

\section{The non-collinear Ising-Glauber model}
In a celebrated paper~\cite{Glauber}, Glauber introduced, in the
usual 1D Ising model~\cite{Ising}, a stochastic dependence on the
time variable $t$: \textit{i.e.}, the state of a spin lying on the $k$-th
lattice site was represented by a two-valued stochastic function
$\sigma_{k}(t)$
\begin{equation}
\label{Ising_Hamiltonian_in_Glauber} \mathcal{H}_{I}=-\sum
^{N}_{k=1}\, \left( J_I \sigma_{k}\sigma_{k+1}+ g \mu_B {\rm
H}e^{-i\omega t} \sigma_k \right),~~\sigma_{k}(t)= \pm 1.
\end{equation}
$J_I$ is the exchange coupling constant, that favors nearest
neighboring spins to lie parallel ($J_I>0$, ferromagnetic
exchange) or antiparallel ($J_I<0$, antiferromagnetic exchange);
$g$ is
the Land\'e factor of each spin, and $\mu_B$ the Bohr magneton.
In the original paper~\cite{Glauber} a 1D lattice
of equivalent and collinear spins was studied; there the response
to a time-dependent magnetic field H$(t)$, applied parallel to the
axis of spin quantization and oscillating with frequency $\omega$,
as in typical {\it a.c.} susceptibility experiments, was also considered.

In order to investigate the phenomena of slow relaxation (for H=0)
and resonant behavior of the {\it a.c.} susceptibility (for H$\ne
0$) in molecular SCM's, we are going to generalize
the Glauber model (\ref{Ising_Hamiltonian_in_Glauber}) accounting
for non-collinearity of local anisotropy axes and crystallographic
(laboratory) frame. To this aim, we adopt the following model
Hamiltonian
\begin{equation}
\label{non_collinear_H} \mathcal{H}=-\sum ^{N}_{k=1}\, \left( J_I
\sigma_{k}\sigma_{k+1} +G_k \mu_B {\rm H}e^{-i\omega t}\sigma_{k}
\right),~~\sigma_{k}(t)= \pm 1.
\end{equation}
$J_I$ is an effective Ising exchange coupling that can approximately
be related to the Hamiltonian parameters of a real
SCM~\cite{Ale_Clerac_2007,Ale_per_Gatto_2008}: see
later on the discussion in Section V.
Like in the usual Ising-Glauber
collinear model (\ref{Ising_Hamiltonian_in_Glauber}), the spins in
Eq.~(\ref{non_collinear_H}) are described by classical, one-component vectors
that are allowed to take two integer values $\sigma_{k}(t)=\pm 1$,
but now the magnetic moments
may be oriented along different directions, $\hat{\mathbf{z}}_{k}$,
varying from site to site.
Within this scheme, the Land\'e tensor of a spin on the
$k$-th lattice site has just a non-zero component,
$g_k^{\Vert}$, along the local easy anisotropy direction $
\hat{\mathbf{z}}_{k}$. Denoting by $\hat{\mathbf{e}}_{ \mathrm{H}}$ the direction
of the oscilla.ting magnetic field, $\mathbf{H}(t)$=$\mathrm{H} e^{-i\omega
t} \hat{\mathbf{e}}_{ \mathrm{H}}$, we define the
generalized Land\'e factor $G_k$ appearing in Eq. (\ref{non_collinear_H})
as \begin{equation}
\label{G_k_def}
G_k =
g_k^{\Vert}~\hat{\mathbf{z}}_{k}\cdot\hat{\mathbf{e}}_{\mathrm{H}}
\end{equation}
\textit{i.e.}, a scalar quantity that varies from site to site.
Following Glauber~\cite{Glauber}, we define
the single spin expectation value
$s_{k}(t)=\left\langle \sigma_{k}\right\rangle_t$, where brackets denote
a proper ensemble average,  and the
stochastic magnetization along the direction of the applied field
\begin{equation} \label{Stochastic_Mag_Glauber}
 \left\langle M \right\rangle_t =  \mu_{B}\sum
^{N}_{k=1} G_k \left\langle \sigma_{k}\right\rangle_t=\mu_{B}\sum
^{N}_{k=1} G_k s_{k}(t)\,.
\end{equation}
The basic equation of motion of the Glauber model~\cite{Brey_Prados,da_Silva}  reads
\begin{equation} \label{Glauber_eq_Motion}
\frac{d}{dt}s_{k}(t)=
-2\left\langle\sigma _{k}w_{\sigma
_{k}\rightarrow -\sigma _{k}}\right\rangle\,,
\end{equation}
in which $ w_{\sigma _{k}\rightarrow -\sigma _{k}}$ represents the
probability per unit time to reverse the $k$-th spin, through the
flip +$\sigma _{k}\rightarrow -\sigma _{k}$. For a system of $N$
coupled spins, this probability is affected by the interaction
with the other spins, with the thermal bath and, possibly, with an
external magnetic field. Among all possible assumptions for the
transition probability $w_{\sigma_{k}\rightarrow -\sigma_{k}}$ as
a function of the $N+1$
variables~\cite{Kimball,Felderhof_Suzuki,Einax_Schulz,Berim_Ruckenstein}
 $\{\sigma _{1},...,\sigma _{k},...,\sigma _{N},t\} $, again following Glauber~\cite{Glauber} we require
$w_{\sigma_{k}\rightarrow -\sigma_{k}}$ to be independent of time
and to depend only on the configuration of the two nearest
neighbors of the $k$-th spin. In zero field, these requirements
are fulfilled by
\begin{equation}
\label{Transition_Prob_Glauber} w^{\rm H=0}_{\sigma
_{k}\rightarrow -\sigma _{k}}=\frac{1}{2}\alpha\left[
1-\frac{1}{2}\gamma \sigma_{k}\left(
\sigma_{k-1}+\sigma_{k+1}\right) \right]
\end{equation}
while in the presence of an external field
\begin{equation}
\label{Transition_Prob_Field_Glauber} w^{ \mathrm{H}}_{\sigma
_{k}\rightarrow -\sigma _{k}}= w^{\rm H=0}_{\sigma _{k}\rightarrow
-\sigma _{k}}~\left( 1-\delta_k \sigma _{k}\right)\,,
\end{equation}
is usually chosen; the attempt frequency $\frac{1}{2}\alpha$ (\textit{i.e.}, the probability per unit
time to reverse an isolated spin) remains an undetermined
parameter of the model; $\gamma$ accounts for the effect of the
nearest neighbors; the parameters $\delta_k$ have the
role of stabilizing the configuration in which the $k$-th spin is
parallel to the field, and destabilizing the antiparallel
configuration.
Thanks to the particular choices \eqref{Transition_Prob_Glauber}
and   \eqref{Transition_Prob_Field_Glauber} for the transition
probability, by imposing the Detailed Balance
conditions~\cite{Glauber} it is possible to express $\gamma$ and
$\delta_k $ as functions of the parameters in the spin Hamiltonian (\ref{non_collinear_H})
\begin{equation}
 \label{gamma_Glauber} \gamma =\tanh \left(2\beta J_I\right)
,~~~ \delta_k =\tanh \left( \beta G_k\mu_{B} \mathrm{H}\right)\,.
\end{equation}
where $\beta={1\over {k_B T}}$ is the inverse temperature in units
of Boltzmann's constant. 
Another advantage of Glauber's choices
\eqref{Transition_Prob_Glauber} and
\eqref{Transition_Prob_Field_Glauber} is that the equation of
motion \eqref{Glauber_eq_Motion} takes a simple form. In
particular, for ${\rm H}=0$, Eq.~(\ref{Glauber_eq_Motion}) with
the choice (\ref{Transition_Prob_Glauber}) becomes
\begin{equation} \label{Glauber_eq_Motion_Mat_Ring}
\frac{d\underline{s}(t)}{dt}=- \alpha\mathbf{A}\underline{s}(t)
\,,
\end{equation}
\noindent where $\underline{s}(t)$ denotes the vector of single-spin
expectation values $\{s_1(t),~s_2(t),\cdots,~s_N(t)\}$ and
$\mathbf{A}$ is a square $N \times N$ symmetric matrix, whose
non-zero elements are A$_{k,k}$=1 and
A$_{k,k-1}$=A$_{k,k+1}$=$-\frac{\gamma}{2}$, with
A$_{1,N}$=A$_{N,1}$=$-\frac{\gamma}{2}$ if periodic boundary
conditions are assumed for the $N$-spin chain. A closed solution
of this set of first-order differential equations can be obtained
expressing the expectation value of each spin, $s_k(t)$, in terms
of its spatial Fourier Transform (FT) $\widetilde{s}_q$
\begin{equation} \label{xxx}
s_k(t)=\sum_q \widetilde{s}_q e^{iqk}e^{-\lambda_q t}.
\end{equation}
\noindent Substituting Eq.~(\ref{xxx}) into
Eq.~(\ref{Glauber_eq_Motion_Mat_Ring}), one readily obtains the
dispersion relation
\begin{equation} \label{Glauber_dispersion_pbc}
\lambda_q = \alpha\left(1- \gamma \cos q \right),~~~q=\frac{2
\pi}{N}n
\end{equation}
\noindent with $n=0,~1,...,~N-1$~\cite{Suzuki_Kubo}. For
ferromagnetic coupling ($J_I>0$, hence $\gamma>0$) the smallest
eigenvalue $\lambda_{q=0}=\alpha(1-\gamma)$ occurs for $n$=0,
independently of the number of spins $N$ in the chain. For
antiferromagnetic coupling ($J_I<0$, hence $\gamma<0$) and $N$
even, the smallest eigenvalue $\lambda_{q=\pi}=\alpha(1-|\gamma|)$
occurs for $n$=$\frac{N}{2}$; while in the case of $N$ odd, the
smallest eigenvalue corresponds to
$\alpha\left[1-|\gamma|\cos\left( \frac{\pi}{N} \right)\right]$,
thus depending on the number of spins in the antiferromagnetic
chain~\cite{Luscombe}.  The characteristic time scales of the
system, $\tau_q$, are given by
\begin{equation}
\label{tauq} \tau_q={1\over {\lambda_q}}={1\over {\alpha(1-\gamma
\cos q)}}.
\end{equation}
At finite temperatures the characteristic times $\tau_q$ are
finite because $|\gamma|<1$; for $T \to 0$ one has that
$1-|\gamma|$ vanishes irrespectively of the sign of $J_I$, because
$\gamma \to {{J_I} \over {\vert J_I \vert}}=\pm 1$. Thus, for
H$=0$, there is one diverging time scale in the $T \to 0$ limit:
$\tau_{q=0}$ for ferromagnetic coupling and $\tau_{q=\pi}$ for
antiferromagnetic coupling (and even $N$). In the presence of a
non-zero, oscillating field ${\rm H}(t)={\rm H} e^{-i\omega t}$,
the equation of motion~(\ref{Glauber_eq_Motion}) with the choice
(\ref{Transition_Prob_Field_Glauber}) takes a form (see
Eq.~\ref{Glauber_eq_Motion_Field_Ring} in Section IV later on)
which can still be solved, though in an approximate
way~\cite{Glauber}, for a sufficiently weak intensity of the
applied magnetic field.

\section{Relaxation of the magnetization in zero field}
The original Glauber model was formulated for a chain of collinear spins
with the same Land\'e factors: {\it i.e.}, in
Eq.~(\ref{non_collinear_H}) one has $G_k=g$, $\forall k=1,\cdots,N$.
Assuming that the system has been fully magnetized by means of a
strong external field, one can study how the system evolves if the field is
removed abruptly. This corresponds to take a fully saturated initial
condition \begin{equation}
\label{sat} s_k(0)=1,~~~\forall k.
\end{equation}

In  ferrimagnetic chains, on the other hand, a ``partial"
saturation can be reached, provided the antiferromagnetic coupling
($J_I<0$) between nearest neighbors is ``strong enough", in a
sense that will be clarified later on. In fact, if the Land\'e
factors for odd  and even sites are not equal ($g_{o} \ne g_{e}$),
through the application of an opportune field the sample can be
prepared in a configuration with
\begin{equation}
\label{partial_sat}
\begin{cases} s_k(0)=+1,~~ {\rm for}~~k=2r+1~~(k~~ {\rm
odd})\\s_k(0)=-1,~~{\rm for}~~k=2r~~(k~~{\rm even}).
\end{cases}
\end{equation}
With respect to the case considered by Glauber, it is convenient
to separate explicitly the expectation values of the odd
sites, $s_{2r+1}(t)$, from those of the even sites, $ s_{2r}(t)$.
Thus, for ${\rm H}=0$, the set of $N$ equations of motion
(\ref{Glauber_eq_Motion_Mat_Ring}) can be rewritten as
\begin{equation}
\label{newform}
\begin{cases}
\frac{d}{dt}s_{2r}=-\alpha \left[s_{2r}+\frac{1}{2}\gamma (s_{2r+1}+s_{2r-1})\right]\\
\frac{d}{dt}s_{2r+1}=-\alpha \left[s_{2r+1}+\frac{1}{2}\gamma
(s_{2r}+s_{2r-2})\right]
\end{cases}
\end{equation}
In the following, the solutions of (\ref{newform}) will be
found using two different approaches that yield identical results.

\subsection{The Generating Function approach}
The Generating Function approach, which closely follows the
original Glauber's paper, is exposed in detail in Appendix.
Here, in order to distinguish between the ferromagnetic and
ferrimagnetic relaxations, we specialize the general solution, Eq.
\eqref{odd_sites_relax_Solution} and Eq. \eqref{even_sites_relax_Solution},
to the two different kinds of initial
conditions, Eq.~(\ref{sat}) and Eq.~(\ref{partial_sat}). In both
cases, we will assume that the exchange coupling $J_I$ is
negative. The ``partially saturated'' configuration,
Eq.~(\ref{partial_sat}), reflects a typical experimental
situation, in which the antiferromagnetic coupling is much bigger
($J_I \approx$ 100$\div$1000 K) than the Zeeman energy associated
with accessible magnetic fields.  On the other hand, the initial
configuration with all the spins aligned in the same direction,
Eq.~(\ref{sat}), clearly reflects the experimental situation of a
fully saturated sample. This condition is easily obtained for
ferromagnetic coupling ($J_I>0$), while it may require very strong
fields (eventually unaccessible) for antiferromagnetic coupling
($J_I<0$).

Let us start from the saturated configuration, Eq.~(\ref{sat}).
Substituting the initial condition $s_k(0)=1$ for all $k$ in both
\eqref{odd_sites_relax_Solution} and
\eqref{even_sites_relax_Solution}, we obtain
\[ \begin{cases} s_{2r}(t)=e^{-\alpha t}
\overset{+\infty }{\underset{m=-\infty}\sum} \left[
\mathcal{I}_{2(r-m)}(\gamma \alpha t)+\mathcal{I}_{2(r-m)-1}(\gamma
\alpha t)\right]
\\
s_{2r+1}(t)=e^{-\alpha t}\overset{+\infty
}{\underset{m=-\infty}\sum} \left[ \mathcal{I}_{2(r-m)}(\gamma
\alpha t)+\mathcal{I}_{2(r-m)+1}(\gamma \alpha t)\right].
\end{cases} \]
Hence, exploiting the property  \eqref{Bessel_Prop} of the Bessel
functions (taking $y =1$), and redefining the sums
by a unique index $j$, we get \begin{equation}
\begin{cases}
\label{ferro_sites_relax_Solution} s_{2r}(t)=e^{-\alpha t}
\overset{+\infty }{\underset{j=-\infty}\sum}
 \mathcal{I}_{j}(\gamma \alpha t)=e^{-\alpha (1-\gamma )t}\\
s_{2r+1}(t)=e^{-\alpha t}\overset{+\infty
}{\underset{j=-\infty}\sum} \mathcal{I}_{j}(\gamma \alpha
t)=e^{-\alpha (1-\gamma )t}.
\end{cases}\end{equation}
This means that, starting with all the spins aligned in the same
direction, each spin expectation value (both on even and odd
sites) decays obeying a mono-exponential law with relaxation time
$\tau_{q=0}=\left[\alpha (1-\gamma)\right]^{-1}$, which is just
the characteristic time scale obtained as the inverse of the
dispersion relation $\lambda_{q}$ with zero wave number $q=0$, see
Eq. \eqref{Glauber_dispersion_pbc}. Notice that $\tau_{q=0}$ can
diverge for $T \to 0$ only in the case of ferromagnetic coupling,
$J_I>0$ ($\gamma>0$).

Let us now consider the partially saturated configuration,
Eq.~(\ref{partial_sat}), in which $s_k(0)=1$ for $k$ odd and
$s_k(0)=-1$ for $k$ even. Substituting these initial conditions in
both \eqref{odd_sites_relax_Solution} and
\eqref{even_sites_relax_Solution}, we obtain
\[\begin{cases}
s_{2r}(t)= -e^{-\alpha t}\overset{+\infty
}{\underset{m=-\infty}\sum} \left[ \mathcal{I}_{2(r-m)}(\gamma
\alpha t)-\mathcal{I}_{2(r-m)-1}
(\gamma \alpha t)\right] \\
s_{2r+1}(t)= e^{-\alpha t}\overset{+\infty
}{\underset{m=-\infty}\sum} \left[ \mathcal{I}_{2(r-m)}(\gamma
\alpha t)-\mathcal{I}_{2(r-m)+1}(\gamma \alpha t)\right]
\end{cases} \]
and, still exploiting the property \eqref{Bessel_Prop} (but now
for $y =-1$), we get
\begin{equation}
\begin{cases}
\label{ferri_sites_relax_Solution} s_{2r}(t)=-e^{-\alpha
t}\overset{+\infty }{\underset{j=-\infty}\sum}(-1)^{j}
 \mathcal{I}_{j}(\gamma \alpha t)=-e^{-\alpha (1+\gamma )t}\\
s_{2r+1}(t)=e^{-\alpha t}\overset{+\infty
}{\underset{j=-\infty}\sum}(-1)^{j} \mathcal{I}_{j}(\gamma \alpha
t)=e^{-\alpha (1+\gamma )t}\,.
\end{cases}\end{equation}
Also in this case all the spins of the system relax with a
mono-exponential law, but now the relaxation time is 
$\tau_{q=\pi}=\left[\alpha (1+\gamma )\right]^{-1}$,  which
corresponds to the inverse of the eigenvalue $\lambda_q$
with wave number $q=\pi$, see Eq.~(\ref{Glauber_dispersion_pbc}).
Notice that $\tau_{q=\pi}$ can diverge for $T \to 0$ only in the
case of antiferromagnetic coupling, $J_I<0$ ($\gamma<0$).

Summarizing, according to the sign of the exchange constant, both
time scales $\tau_{q=0}$ (for $J_I>0$) and $\tau_{q=\pi}$ (for
$J_I<0$) diverge in the low temperature limit $T \to 0$, following
an Arrhenius law
\begin{equation}
\label{tau_divergence_infty} \tau =\frac{1}{2\alpha} e^{4\beta|J_I|}
\end{equation}
with energy barrier $4|J_I|$ (slow relaxing mode). It is worth
noting that the remaining relaxation times, given by the inverse
of the eigenvalues in Eq.~\eqref{Glauber_dispersion_pbc} with
$q\neq 0$ and $q\ne\pi$, always remain of the same order of
magnitude as $\alpha^{-1}$ (fast relaxing modes). This time scale
is typically very small ($\sim$ ps) in real systems, and
negligible with respect to the characteristic times involved in any
experimental measurement we refer to.

\subsection{The Fourier Transform approach}
The solutions, \eqref{ferro_sites_relax_Solution} and
\eqref{ferri_sites_relax_Solution}, to the set of equations
(\ref{newform}) can alternatively be deduced within the Fourier
Transform (FT) formalism, which has already been exploited to obtain
the dispersion relation \eqref{Glauber_dispersion_pbc}. 
Recalling the definition \eqref{xxx} of $s_k(t)$ and its
spatial FT
\begin{equation}
\label{yyy}
\widetilde{s}_{q}=\frac{1}{N} \sum_k s_k(t)
e^{-iqk}e^{\lambda_{q} t},
\end{equation}
\noindent we evaluate $\widetilde{s}_{q}$ at time $t=0$,
$\widetilde{s}_{q}=\frac{1}{N} \sum_k s_k(0) e^{-iqk}$, for the
two initial conditions of interest, (\ref{sat}) and (\ref{partial_sat}).
Starting from the all-spin-up configuration, Eq.~(\ref{sat}), we have
\begin{equation}
\widetilde{s}_{q}=\frac{1}{N} \sum_k  e^{-iqk}=\delta_{q, 0}\,.
\end{equation}
\noindent Hence the solution for the expectation value of a spin
localized on the $k$ lattice site at time $t$ is
\begin{equation}
\label{ferro_sites_relax_Solution_FT} s_k(t)=\sum_q \delta_{q, 0}
e^{iqk} e^{-\lambda_q t}=e^{-\lambda_{q=0}\, t}\,,
\end{equation}
\noindent which is identical to \eqref{ferro_sites_relax_Solution}
since $\lambda_{q=0} =\alpha(1-\gamma)$.

Starting from the partially saturated configuration,
Eq.~(\ref{partial_sat}), it is useful to rewrite it as
$s_k(0)=-e^{i\pi k}$, so that the FT at $t=0$ is
\begin{equation}
\widetilde{s}_{q}=-\frac{1}{N} \sum_k  e^{i\pi k}
e^{-iqk}=-\delta_{q, \pi}\,.
\end{equation}
Hence the solution is readily obtained
\begin{equation}
\label{ferri_sites_relax_Solution_FT} s_k(t)=-\sum_q \delta_{q,
\pi} e^{iqk} e^{-\lambda_q t}=-e^{i\pi k} e^{-\lambda_{q=\pi} t},
\end{equation}
\noindent which is identical to \eqref{ferri_sites_relax_Solution}
since $\lambda_{q=\pi} =\alpha(1+\gamma)$.

Finally, we observe that Eqs. \eqref{ferro_sites_relax_Solution_FT}
and \eqref{ferri_sites_relax_Solution_FT} hold even for a ring
with a {\it finite} number $N$ of spins, while Eqs.
\eqref{ferro_sites_relax_Solution} and
\eqref{ferri_sites_relax_Solution} were obtained in the
infinite-chain limit.

\subsection{Slow versus fast relaxation of the spontaneous magnetization}
The expectation values, $s_k(t)$, of spins localized on the even
and odd sites of a linear lattice at time $t$, computed either
with the Generating Functions or the Fourier Transform approach,
have been shown to display a mono-exponential relaxation, see
Eqs.~\eqref{ferro_sites_relax_Solution} and
\eqref{ferri_sites_relax_Solution}, with different time scales,
$\tau_{q=0}=[\alpha (1-\gamma)]^{-1}$ and $\tau_{q=\pi}[\alpha (1+\gamma)]^{-1}$
respectively, depending on the different initial conditions,
Eqs.~(\ref{sat}) and (\ref{partial_sat}). As a consequence, also
the macroscopic magnetization, expressed  by
Eq.~\eqref{Stochastic_Mag_Glauber}, displays the same
mono-exponential relaxation as the single site quantities
$s_k(t)$.

A chain in which all the magnetic moments are equal can be
prepared only in the saturated initial configuration,
Eq.~(\ref{sat}), with all the spin aligned in the same direction,
through the application of an external field. Thus, when the field
is abruptly removed, such a system will relax slowly at low
temperature only if the exchange coupling is ferromagnetic
($J_I>0$). In contrast, if the exchange coupling is
antiferromagnetic ($J_I<0$) and the chain is ``forced" in the
saturated state by a strong applied magnetic field, the system
will relax very fast (in a typical time of the order of
$\alpha^{-1}$) when the field is removed. Let us now discuss how
these results, first obtained by Glauber~\cite{Glauber}, are
generalized to the case of a chain in which the magnetic moments
are collinear, but {\it not} equal on each site.

As pointed out in the introduction, a model with antiferromagnetic
coupling ($J_I<0$) but non-compensated magnetic moments on the two
sublattices is more akin to real SCM's~\cite{Review_CCM,feature_SCM,Caneschi_Angew}.
Yet it is very interesting since, depending on the intensity of the
applied magnetic field, the system can be prepared either in the
saturated initial configuration, Eq.~(\ref{sat}), where all spins
are parallel to each other, or in the partially saturated one
Eq.~(\ref{partial_sat}), where nearest neighbors are antiparallel.
In the former case, a very strong field is required in order to
overcome the antiferromagnetic coupling between nearest neighbors;
once the field is removed, the relaxation of the magnetization is
expected to be fast at low temperatures, on the basis of the
solution \eqref{ferro_sites_relax_Solution}. In the latter case,
the partially saturated initial configuration (\ref{partial_sat})
can easily be obtained through the application of a smaller,
experimentally accessible magnetic field; when the field is
abruptly removed, the relaxation is expected to be slow according
to the solution \eqref{ferri_sites_relax_Solution}. The solution
\eqref{ferri_sites_relax_Solution} justifies the observation of SCM behavior in
ferrimagnetic quasi-1D compounds like CoPhOMe~\cite{Caneschi_Angew}
(see Sect. V).

Summarizing, we have found that when a collinear ferrimagnetic
chain is prepared in an initial state --  fully or partially
saturated depending on the intensity of the applied magnetic field 
-- once the field is removed abruptly, the spin system can show
fast or slow relaxation, respectively. Fast relaxation corresponds
to stronger fields; unfortunately for the quasi-1D chain compound
CoPhOMe~\cite{Caneschi_Angew,CoPhOMe_EPL}, the antiferromagnetic
exchange constant is so large ($\vert J_I \vert \sim 100$K)
that the realization of a fully saturated initial configuration
 would require a very high, almost unaccessible field
($\sim 1000$ kOe). Thus this compound is not a good candidate for
such a kind of experiments~\cite{nota_equil}.

\section{Magnetic response to an oscillating  magnetic field}

\begin{figure*}\label{collinear}
\includegraphics[width=16cm,angle=0,bbllx=56pt,bblly=280pt,%
bburx=545pt,bbury=587pt,clip=true]{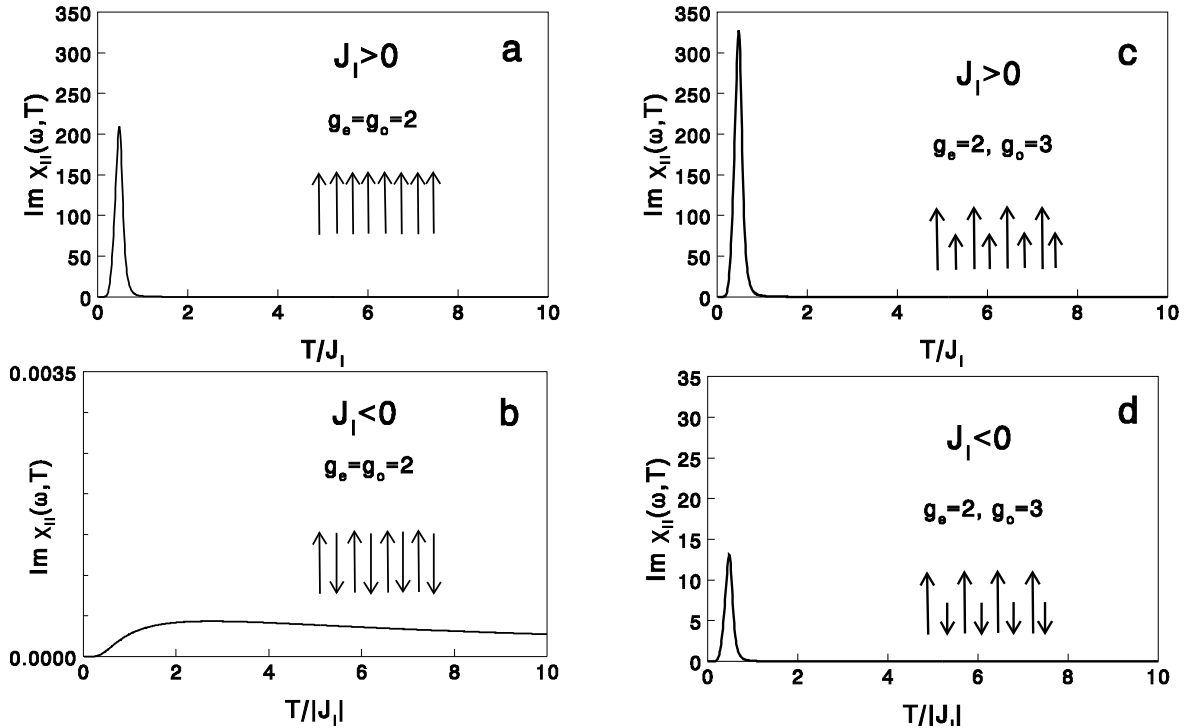}
\caption{Temperature dependence of the imaginary part of the
complex susceptibility, Eq.~
(\ref{ac_Susceptibility_Ferrimagnet_Glauber}), of a collinear
one-dimensional Ising model with alternating spins. Resonant
behavior in response to an oscillating magnetic field is possible,
at low frequency, only when magnetic moments are uncompensated (a,c,d),
while a broad peak is found when the net magnetization is zero (b).
(The curves refer to reduced frequency $\omega/\alpha=0.001$)}
\end{figure*}

In the presence of a magnetic field H, the transition probability
to be put in the equation of motion  \eqref{Glauber_eq_Motion} is
$w^{ \mathrm{H}}_{\sigma _{k}\rightarrow -\sigma _{k}}$, defined
in Eq.~\eqref{Transition_Prob_Field_Glauber}. One obtains
\begin{eqnarray} \label{Glauber_eq_Motion_Field_Ring}
 \frac{ds_k(t)}{d t}&=&-\alpha \Big\{ s_k(t) - \frac{\gamma}{2}\left[
s_{k+1}(t)+s_{k-1}(t)\right] \cr &+&\frac{\gamma\delta_k}{2}\left[
\left\langle \sigma_{k}\sigma_{k+1}\right\rangle_t + \left\langle
\sigma_{k-1}\sigma_{k}\right\rangle_t \right] -\delta_k \Big\}
\end{eqnarray}
that differs from Eq.~(\ref{newform}), considered earlier for
H$=0$, in the presence of both a non-homogeneous term, $\delta_k$, and
the time-dependent pair-correlation functions
$\left\langle \sigma_{k}\sigma_{k\pm 1}\right\rangle_t$.
The latter ones, assuming that the field is so weak to induce just small
departures from equilibrium, can be approximated by their
time-independent counterparts~\cite{Huang}
\begin{equation}
\left\langle \sigma_{k}\sigma_{k+1}\right\rangle_t= \left\langle
\sigma_{k-1}\sigma_{k}\right\rangle_t \approx \tanh(\beta
J_I)\equiv \eta.
\end{equation}
As it is usual in  \textit{a.c.} susceptibility measurements, we
also assume the time-dependent magnetic field
$\mathbf{H}(t)$=$\mathrm{H} e^{-i\omega t} \hat{\mathbf{e}}_{
\mathrm{H}}$, oscillating at frequency $\omega$, to be weak so
that the $\delta_k$ parameters can be linearized
\begin{equation}
\label{delta_Glauber_linear} \delta_k =\tanh \left( \beta \mu_{B}
G_k \mathrm{H}(t)\right) \approx \beta \mu_{B}
G_k \mathrm{H}(t).
\end{equation}
The system of equations of motion
\eqref{Glauber_eq_Motion_Field_Ring} then takes the form
\begin{eqnarray} \label{Ring_eq_motion_geometry}
\frac{ds_{k}(t)}{d(\alpha t)}&=&-s_{k}(t)+\frac{\gamma }{2}\left[
s_{k+1}(t)+s_{k-1}(t)\right]\cr &+&  \beta  f(\beta J_I) \mu_{B}
G_k \mathrm{H}(t)\,,
\end{eqnarray}
where
\begin{equation}
f(\beta J_I)=1-\gamma \eta={{1-\eta^2}\over {1+\eta^2}}
\end{equation}
is a function of the reduced coupling constant $\beta J_I$
and we have taken into account that $\gamma=2\eta/(1+\eta^2)$. After a
brief transient period, the system will reach the stationary
condition in which the magnetic moment of each spin oscillates
coherently with the forcing term at the frequency $\omega$. Expressing the
expectation value of a spin on the $k$-th lattice site, $s_k(t)$,
through its spatial FT, $\widetilde{s}_q$, the
trial solution is
\begin{equation}
s_k(t)=\sum_q \widetilde{s}_q e^{iqk}e^{-i \omega t}\,.
\end{equation}
Substituting the latter in the system
\eqref{Ring_eq_motion_geometry}  we get
\begin{equation}
\label{new} \widetilde{s}_q = \beta f(\beta J_I) \mu_B \mathrm{H}
{{\alpha \widetilde{G}_q}\over {\alpha(1-\gamma \cos q)-i\omega}},
\end{equation}
where $\widetilde{G}_q$ is the FT of $G_k$:
\begin{equation}
\widetilde{G}_q={1\over N}\sum_{k=1}^N e^{-iqk} G_k.
\end{equation}
The average of stochastic magnetization can readily be obtained
from (\ref{Stochastic_Mag_Glauber}) as
\begin{equation} \label{Stochastic_Mag_ac_Glauber_one}
 \left\langle M \right\rangle_t = \mu_{B} e^{-i \omega t} \sum^{N}_{k=1}
\sum_{q q'} \widetilde{G}_q  \widetilde{s}_{q'} e^{iqk}  e^{iq' k}
\,,
\end{equation}
which accounts for non-collinearity of local anisotropy axes with
respect to the field direction. Performing the sum over all the
lattice sites ($k$ indices) yields a factor $N\delta_{q, -q'}$ in
Eq. \eqref{Stochastic_Mag_ac_Glauber_one}; substituting the
expression (\ref{new}) for $\widetilde{s}_q$, one obtains
\begin{equation} \label{Stochastic_Mag_ac_Glauber_two}
 \left\langle M \right\rangle_t = N \mu^2_{B} \beta f(\beta
J_I) \mathrm{H} e^{-i \omega t} \sum_{q q'} \frac{\alpha
\widetilde{G}_q \widetilde{G}_{q'} \delta_{q, -q'} }{\alpha
(1-\gamma \cos q ) -i\omega} \,.
\end{equation}
Then, considering that $\widetilde{G}_q
\widetilde{G}_{-q}=|\widetilde{G}_q |^2$, the \textit{a.c.}
susceptibility is finally obtained dividing
\eqref{Stochastic_Mag_ac_Glauber_two} by $ \mathrm{H}e^{-i \omega
t}$
\begin{equation} \label{ac_Susceptibility_geom_Glauber}
 \chi(\omega, T) = N \mu^2_{B} \beta f(\beta
J_I) \sum_{q} \frac{ \alpha|\widetilde{G}_q |^2}{\alpha (1-\gamma
\cos q ) -i \omega} \,.
\end{equation}
In principle, the {\it a.c.} susceptibility of a chain with $N$
spins admits $N$ poles, corresponding to the $N$ eigenvalues
$\lambda_q$ in Eq.~\eqref{Glauber_dispersion_pbc}. Each mode is
related to a different time scale $\tau_q=1/\lambda_q$. In
practice, not all the time scales will be involved in the complex
susceptibility  $\chi(\omega, T)$, but {\it only the ones selected
by} $\widetilde{G}_q$. A result similar, at first glance, to
Eq.~(\ref{ac_Susceptibility_geom_Glauber}) was deduced by Suzuki
and Kubo~\cite{Suzuki_Kubo}, but in their case the relationship
was between the time scale $\tau_q$ and the wave-vector-dependent
susceptibility $\chi(q, \omega)$. In contrast, in an \textit{a.c.}
susceptibility experiment only the zero-wave-vector susceptibility
$\chi(q=0,\omega)$ is accessible; the peculiarity of
Eq.~(\ref{ac_Susceptibility_geom_Glauber}) is that other time
scales, different from $\tau_{q=0}$, can be selected thanks to the
dependence of the gyromagnetic factors $G_k$ and of the local anisotropy
axes on the site position $k$. This is the main result of our study and
will be clarified hereafter through a few examples.

\subsection{The \textit{a.c.} susceptibility of a collinear Ising ferrimagnetic chain}
Let us start considering the case of a one-dimensional Ising model
with two kinds of spins (aligned parallel on antiparallel to the
chain axis) alternating on the odd and even magnetic sites of the
lattice with Land\'e factors $G_{2r+1}=g_{o}$ and $G_{2r}=g_{e}$
(integer $r$), respectively. Strictly speaking, a collinear Ising
ferrimagnet is characterized by an antiferromagnetic coupling
$J_I<0$, but also the case $J_I>0$ can be treated through Eq.
(\ref{ac_Susceptibility_geom_Glauber}). In fact, since the local
axis of anisotropy has the same direction for all the spins, the
FT of the site-dependent Land\'e factor is
\begin{eqnarray}
\label{collinear_alt} \widetilde{G}_q&=& {1\over
{N}}\sum_{r=1}^{N/2} [ g_e e^{-iq 2r}+g_o e^{-i q(2r-1)}]
\cr&=&(g_{e}+e^{iq}g_{o} ){1\over {N}}\sum_{r=1}^{N/2} e^{-iq
2r}.\end{eqnarray}
Taking into account that, in the presence of
periodic boundary conditions, one has
\begin{equation} \label{boundary} \sum_{r=1}^{N/2} e^{-iq 2r}= {{N}\over
2}(\delta_{q,0}+\delta_{q,\pi}),\end{equation} it follows that the
only non-zero values of $\widetilde{G}_q$ are for $q=0$ and $q=\pi$
\begin{equation}
\label{collinear_altern}
 \widetilde{G}_q^{\Vert}={1\over
 2}\left[(g_e+g_o)\delta_{q,0}+(g_e-g_o)\delta_{q,\pi}\right].
 \end{equation}
Thus, according to Eq.~\eqref{ac_Susceptibility_geom_Glauber}, the
parallel {\it a.c.} susceptibility ($\Vert=zz$) of a collinear
Ising chain with alternating spins is
\begin{eqnarray} \label{ac_Susceptibility_Ferrimagnet_Glauber}
\chi_{\Vert}(\omega, T)&=&N \mu^2_{B}\beta~f(\beta J_I)\cr
&\times&{1\over 4}\left\lbrack{{(g_e+g_o)^2}\over {(1-\gamma) -i
({{\omega}\over {\alpha}}) }} +{{ (g_e-g_o)^2}\over {(1+\gamma) -i
({{\omega}\over {\alpha}})}} \right\rbrack\end{eqnarray} It
appears that both the relaxation times obtained by Suzuki and
Kubo~\cite{Suzuki_Kubo} for the ordinary and the staggered
susceptibility  of the usual Ising model, namely
$\tau_{q=0}=[\alpha(1-\gamma)]^{-1}$ and
$\tau_{q=\pi}=[\alpha(1+\gamma)]^{-1}$ respectively, do coexist in
the \textit{a.c.} susceptibility
(\ref{ac_Susceptibility_Ferrimagnet_Glauber}). Notice that, in the
$\omega\rightarrow 0$ limit, the static susceptibility of the
Ising ferrimagnet in zero field~\cite{PiniRettori} is recovered
from Eq.~(\ref{ac_Susceptibility_Ferrimagnet_Glauber}), since one
has $\frac{f(\beta J_I)}{1 \mp \gamma}={{1-\eta^2}\over
{1+\eta^2}}{1\over {1\mp \gamma}}= e^{\pm 2 \beta J_I}$.

As regards the dynamic response of the system to an oscillating
magnetic field applied along the chain axis, depending on the sign
of the effective exchange coupling constant $J_I$, the ferromagnetic
($g_e+g_o$) or the antiferromagnetic ($g_e-g_o$) branch of the
parallel susceptibility
\eqref{ac_Susceptibility_Ferrimagnet_Glauber} are characterized by
a diverging time scale at low temperature. In particular, for a
collinear Ising ferrimagnet one has $J_I<0$, so that
$\tau_{q=\pi}$ is diverging, while $\tau_{q=0}$ is short (of the
order of $\alpha^{-1}$, the attempt frequency of an
isolated spin). Thus, for $J_I<0$, a resonant behavior of the {\it
a.c.} susceptibility versus temperature (at low frequencies
$\omega/\alpha \ll 1$) can only be observed in the case $g_e \ne
g_o$ (see Fig. 1d) when magnetic moments are {\it uncompensated},
while a broad peak is found in the case $g_e=g_o$ when the net
magnetization is zero (see Fig. 1b). Clearly, for $J_I>0$, a
resonant peak is found in both cases (see Fig. 1a and 1c), because
a net magnetization is always present in the system.

Such a resonant behavior of the {\it a.c.} susceptibility versus
$T$, in ferromagnetic~\cite{Brey_Prados_PLA} as well as in
ferrimagnetic~\cite{PiniRettori} Ising chains with single spin-flip
Glauber dynamics, is a manifestation of the stochastic resonance
phenomenon~\cite{Gammaitoni}: \textit{i.e.}, the response of a set of
coupled bistable systems to a periodic drive is enhanced in the
presence of a stochastic noise when a matching occurs between the
fluctuation-induced switching rate of the system and the forcing
frequency. In a magnetic chain, the role of stochastic noise is
played by thermal fluctuations and the resonant peak in the
temperature-dependence of the {\it a.c.} susceptibility occurs when the
statistical time scale, associated to the slow decay of the
magnetization, matches with the deterministic time scale of the
applied magnetic field
\begin{equation}
\label{cond_reson}
\tau_q(T_{peak}) \approx {1\over {\omega}}.
\end{equation}

\subsection{The \textit{a.c.} susceptibility of an $n$-fold helix}
Next, as an example of a non-collinear spin arrangement, we
consider a system of spins with the local axes of anisotropy
arranged on an $n$-fold helix (see Fig. 2); $\theta$ is the angle
that the local axes form with $z$, the unique axis of the helix
(\textit{i.e.}, the chain axis). In this case the Land\'e factors are equal
on all lattice sites, but different spins experience different
fields because of the geometrical arrangement of magnetic moments.
In the following we will make the approximation that the Land\'e
tensor of a spin on the $k$-th lattice site has just a non-zero
component $g$ along the easy anisotropy direction
$\hat{\mathbf{z}}_{k}$, so that $G_k=g \hat{\mathbf{z}}_{k} \cdot
\hat{\mathbf{e}}_{\mathrm{H}}$ (see Eq. \eqref{G_k_def}).
In the crystallographic frame ($x,y,z$),
the directors $\hat{\mathbf{z}}_{k}$ read (integer $k$)
\begin{equation}
\hat{\mathbf{z}}_k=\sin\theta \left[\cos\left({{2\pi k}\over
{n}}\right)\hat{\mathbf{e}}_{x}+\sin\left({{2\pi k}\over {n}}
\right)\hat{\mathbf{e}}_{y}\right]+\cos\theta
\hat{\mathbf{e}}_{z}.\end{equation}

\begin{figure}\label{multihelix}
\includegraphics[width=8cm,angle=0,bbllx=32pt,bblly=212pt,%
bburx=487pt,bbury=547pt,clip=true]{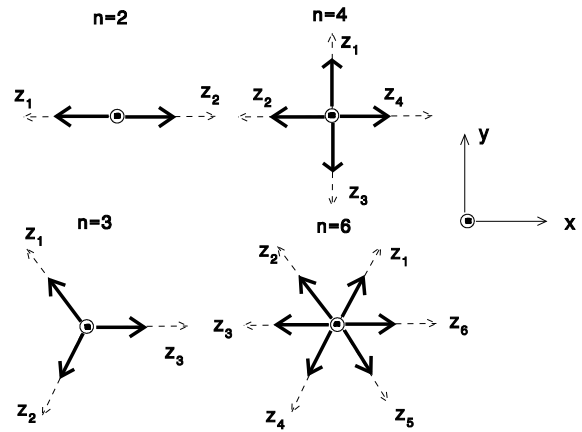} \caption{Thick arrows
denote the projections on the $xy$ plane, perpendicular to the
chain (helix) axis $z$, of magnetic moments in a
one-dimensional Ising helimagnet, for different fold symmetries
($n=2,3,4,6$). Dashed lines are the projections of the local
axes of anisotropy, $\hat{\mathbf{z}}_k$.}
\end{figure}

\begin{figure*}\label{helix}
\includegraphics[width=16cm,angle=0,bbllx=55pt,bblly=270pt,%
bburx=547pt,bbury=572pt,clip=true]{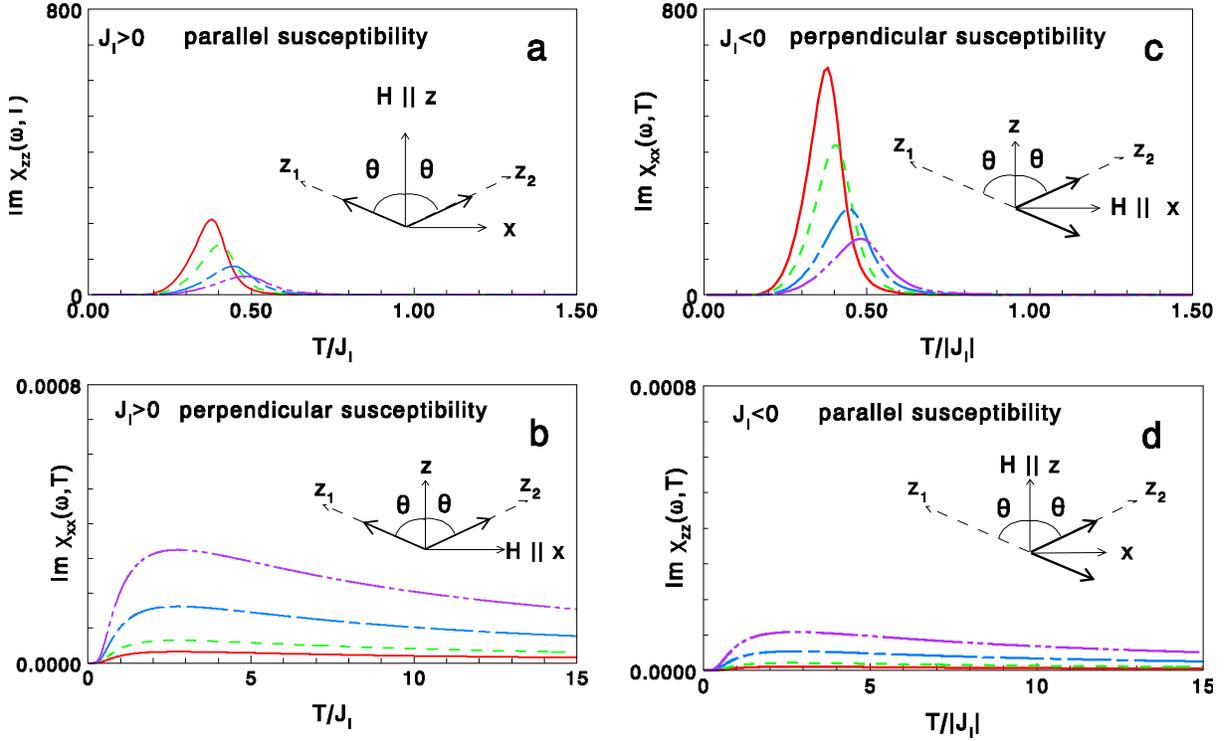} \caption{(color
online) Temperature dependence of the imaginary part of the
parallel (\ref{ac_Susceptibility_zz_helix2_Glauber}) and
perpendicular (\ref{ac_Susceptibility_xx_helix2_Glauber}) complex
susceptibility of an Ising chain with two-fold helical spin
arrangement. The local axes $\hat{\mathbf{z}}_1$ and
$\hat{\mathbf{z}}_2$ were assumed to form an angle $\theta={{\pi}\over 3}$ with
$z$, the chain axis (unique axis of the helix). Different curves
refer to different values of $\omega/\alpha$: 0.0001 (continuous,
red line); 0.0002 (dashed, green line); 0.0005 (dashed
single-dotted, blue line); 0.0010 (dashed double-dotted, violet
line). Resonant behavior in response to an oscillating magnetic
field is possible, at low frequency, only for field applied in a
direction where magnetic moments are uncompensated (a,c), while a broad peak
is found (b,d) when there is no net magnetization along the field direction.}
\end{figure*}

Let us first consider the case of an oscillating magnetic field
H applied parallel to $z$, the helix axis.  All the spins
actually undergo the same field, and since $G_k=g \cos \theta$
independently of the lattice site $k$, the only peak in the
FT $\widetilde{G}_q$ occurs at $q=0$
\begin{equation}
\widetilde{G}_q^{z}=g \cos \theta~
\delta_{q,0}~~(\forall~n).\end{equation}
Following the same
procedure as in the previous paragraph, the parallel {\it a.c.}
susceptibility ($\Vert=zz$) takes an expression (valid for any
value of the fold index $n$ of the helix)
\begin{equation} \label{ac_Susceptibility_zz_helix2_Glauber}
 \chi_{\Vert}(\omega, T) = N \mu^2_{B}\beta~ f(\beta J_I)
 {{g^2 \cos^2\theta}\over {(1-\gamma)-i(
{{\omega}\over {\alpha}})}}~~~(\forall n)
\end{equation}
that differs from Glauber's result for the collinear Ising
chain~\cite{Glauber} only by the geometrical factor
$\cos^2\theta$. For ferromagnetic coupling, $J_I>0$, the
relaxation time $\tau_0=[\alpha(1-\gamma)]^{-1}$ diverges as $T
\to 0$, and a resonant behavior of the {\it a.c.} parallel
susceptibility versus temperature is found, at low frequency, when
the oscillating field is applied parallel to the helix
axis, $z$, along which spins are {\it uncompensated}: see Fig. 3a,
which refers to the case of a two-fold helix ($n=2$) .

Let us now consider the case of an oscillating magnetic field H
applied perpendicularly to the chain axis. In this configuration,
it is useful to distinguish the case $n=2$ from the general case
$n > 2$.

\begin{itemize}

\item $n=2$

In this case, it is worth noticing that for H parallel to $y$,
one has identically $G_r \equiv 0$ for any lattice site $r$. Thus,
$\widetilde{G}_q^y \equiv 0$ and the corresponding {\it a.c.}
susceptibility is identically zero
\begin{equation} \chi_{yy}(\omega, T)\equiv  0\end{equation} (not
shown). In contrast, for H parallel to $x$, one has $G_{r}=-g
\sin \theta$ on odd sites and $G_{r}=+g\sin\theta$ on even sites.
The FT is
\begin{eqnarray} \label{cason2}
\widetilde{G}_q^x&=&{1\over N}\sum_{r=1}^{N/2} g \sin \theta
(-e^{-iq(2r-1)}+e^{-iq2r}) \cr &=&  g\sin\theta {1\over 2}
(\delta_{q,0}+\delta_{q,\pi})(1-e^{iq})\cr&=& g \sin \theta~
\delta_{q,\pi}
\end{eqnarray}
where we have taken into account Eq.~(\ref{boundary}). Thus, for
ferromagnetic coupling, $J_I>0$, the relaxation time
$\tau_{\pi}=[\alpha(1+\gamma)]^{-1}$ does not diverge as $T \to
0$, and the perpendicular {\it a.c.} susceptibility
\begin{equation}
 \label{ac_Susceptibility_xx_helix2_Glauber}
 \chi_{xx}(\omega, T) = N \mu^2_{B} \beta f(\beta J_I)
 {{ g^2 \sin^2\theta }\over {(1+\gamma)-i
({{\omega}\over {\alpha}})}}
\end{equation}
does not present a resonant behavior as a function of temperature;
rather, it presents a broad maximum (see Fig. 3b). Clearly, in the
case of antiferromagnetic coupling, $J_I<0$, the behavior of the
susceptibility components is reversed: a broad maximum is found
for the temperature dependence of the parallel susceptibility
$\chi_{zz}(\omega, T)$ (see Fig. 3d), while a resonant behavior is
found for the perpendicular susceptibility $\chi_{xx}(\omega, T)$
(see Fig. 3c).

\item{$n>2$}

In this case, denoting by  $\hat{\mathbf{e}}_{x}\cdot\hat{\mathbf{e}}_{\mathrm{H}}$ and
$\hat{\mathbf{e}}_{y}\cdot\hat{\mathbf{e}}_{\mathrm{H}}$ the directors of the
in-plane field, the FT's of $G_k$ are given by
\begin{equation}
\begin{cases}
\widetilde{G}^x_q &= {1\over {N}} (\hat{\mathbf{e}}_{x}\cdot\hat{\mathbf{e}}_{\mathrm{H}}) g \sin\theta \sum_{k=1}^{N}
\cos\left(\frac{2\pi k}{n}\right) e^{-iqk} \\
&= (\hat{\mathbf{e}}_{x}\cdot\hat{\mathbf{e}}_{\mathrm{H}})
g \sin\theta \frac{1}{2} \left(\delta_{q, \frac{2\pi}{n} } +  \delta_{q,- \frac{2\pi }{n} } \right)\\
\widetilde{G}^y_q&= {1\over {N}} (\hat{\mathbf{e}}_{y}\cdot\hat{\mathbf{e}}_{\mathrm{H}})  g \sin\theta \sum_{k=1}^{N}
\sin\left(\frac{2\pi k}{n}\right) e^{-iqk} \\
&= (\hat{\mathbf{e}}_{y}\cdot\hat{\mathbf{e}}_{\mathrm{H}}) g \sin\theta
\frac{1}{2 i} \left(\delta_{q, \frac{2\pi}{n} } -  \delta_{q,-
\frac{2\pi }{n} } \right)
\end{cases}.
\end{equation}
Remarkably, as just $|\widetilde{G}_q |^2$ appears in Eq.
\eqref{ac_Susceptibility_geom_Glauber}, the general result for the
in-plane susceptibility turns out to be independent  of the field
direction. Thus, for $n > 2$, the perpendicular ($\perp$) {\it
a.c.} susceptibility of the $n$-fold helix is given by
\begin{eqnarray}
 \label{ac_Susceptibility_perp_helix_Glauber}
&& \chi_{\perp}(\omega, T) =  N \mu^2_{B} \beta~f(\beta J_I)\cr
&\times& \frac{1}{2}\sin^2\theta \frac{ g^2
}{[1-\gamma\cos\left(\frac{2\pi}{n} \right)]
 -i (\frac{\omega}{\alpha})},
\end{eqnarray}
where we have exploited the fact that $\cos\left(-\frac{2\pi}{n}
\right)= \cos\left(\frac{2\pi}{n} \right)$ for the term appearing
at the denominator of Eq. \eqref{ac_Susceptibility_geom_Glauber}.
\end{itemize}

Summarizing, in the general case of an Ising chain with an $n$-fold
helical spin arrangement ($n\ge 2$), we have explicitly shown that
a resonant behavior of the {\it a.c.} susceptibility versus temperature, similar to
the one displayed by ferromagnetic~\cite{Glauber,Brey_Prados_PLA}
and ferrimagnetic~\cite{PiniRettori} Ising chains with collinear
spins, is possible only for field applied in a direction where magnetic moments are
{\it uncompensated}. In contrast, a broad peak is found when there is no net
magnetization along the field direction.

\section{Application to real Single Chain Magnets}
In this Section we will apply the developed formalism to some real
compounds as representative realizations of SCM's;
for the three selected systems
-- we know this restriction is far from being exhaustive~\cite{Review_CCM,feature_SCM} --
\textit{a.c.} susceptibility data on single crystal are available, which is
a fundamental requirement for checking the proposed selection rules.
The considered systems~\cite{Montpellier,tesi_Kevin,Caneschi_Angew} are characterized by 
the alternation of two types of magnetic centers along the chain
axis, so that at least two spins per cell have to be considered;
moreover, the magnetic moments are not collinear, the dominant
exchange interactions are antiferromagnetic and a strong
single-ion anisotropy is present, which favors magnetization
alignment along certain crystallographic directions
$\hat{\mathbf{z}}_{k}$. The static properties of these compounds,
like magnetization and static susceptibility, are generally well
described using a classical Heisenberg model with an isotropic
exchange coupling $J$ and a single-ion anisotropy $D$. Thus, in
order to describe the dynamic behavior in response to a weak,
oscillating magnetic field by means of the previously developed
theory, it is necessary to relate the Hamiltonian parameters of
such a classical spin model to the exchange constant $J_I$ of the
effective Ising model (\ref{non_collinear_H}). In the following we
will show, through a few examples on real systems, that indeed,
depending on the geometry, selection rules are obeyed for the
occurrence of slow relaxation of the magnetization at low
temperatures ($\beta \vert J_I \vert \gg 1$), as well as for
resonant behavior of the {\it a.c.} susceptibility as a function
of temperature at low frequencies. As regards the frequencies
involved in an {\it a.c.} susceptibility experiment on real SCM's,
generally~\cite{Review_CCM,feature_SCM} 
they lie in the range $10^{-1}\div 10^{4}$ Hz, while the attempt
frequency $\alpha$ is of the order of $10^{10}\div10^{13}$ Hz.
Thus, for a typical experiment, a resonant peak in the
\textit{a.c.} susceptibility can safely be observed provided that
at least one of the characteristic time scales $\tau_q$ involved
in \eqref{ac_Susceptibility_geom_Glauber} diverges at low $T$, in
order for the condition (\ref{cond_reson}) to be satisfied.

\subsection{The Mn$^{\rm III}$-based Single Chain Magnet}
In the one-dimensional molecular magnetic compound of formula
[Mn(TPP)O$_2$PPhH]$\cdot$H$_2$O, obtained by reacting Mn(III)
acetate $meso$tetraphenylporphyrin with phenylphosphinic
acid~\cite{Montpellier}, hereafter denoted by Mn$^{\rm III}$-based
SCM, the phenylphosphinate anion transmits a
sizeable antiferromagnetic exchange interaction that, combined
with the easy axis magnetic anisotropy of the Mn$^{\rm III}$
sites, gives rise to a canted antiferromagnetic arrangement of the
spins. The static single-crystal magnetic properties were analyzed
in the framework of a classical spins Hamiltonian
\begin{eqnarray} \label{Heisenberg} \mathcal{H}&=&-\sum
^{N/2}_{r=1}\, \lbrace J {\bf S}_{2r-1}\cdot {\bf S}_{2r} +D
\left\lbrack (S_{2r-1}^{z_{1}})^2 +(S_{2r}^{z_{2}})^2\right\rbrack
\cr &+&e^{-i\omega t} \mu_B {\rm H}^{\alpha}g^{\alpha \beta}\lbrack
S^{\beta}_{2r-1}+S^{\beta}_{2r}\rbrack \rbrace
\end{eqnarray} where $J<0$ is the antiferromagnetic nearest
neighbor exchange interaction between $S=2$ spins. $D>0$ is the
uniaxial anisotropy favoring two different local axes, alternating
along odd and even sites respectively; both axes form an angle
$\theta=21.01^{\rm o}$ with the crystallographic $c$ axis, while
they form opposite angles of modulus $\phi=56.55^{\rm o}$ with the
$a$ axis (see Fig. 4). Thus we can write
$\hat{\mathbf{z}}_{2r-1}=\sin\theta \cos\phi \hat{\mathbf{e}}_{x}
-\sin\theta \sin\phi \hat{\mathbf{e}}_{y}+\cos\theta
\hat{\mathbf{e}}_{z}$ and $\hat{\mathbf{z}}_{2r}=\sin\theta
\cos\phi \hat{\mathbf{e}}_{x} +\sin\theta \sin\phi
\hat{\mathbf{e}}_{y}+\cos\theta \hat{\mathbf{e}}_{z}$.

A best fit of the static single-crystal magnetic susceptibilities,
calculated via a Monte Carlo simulation~\cite{Montpellier}
provides $J=-1.34$ K and $D=4.7$ K; the gyromagnetic tensor
$G^{\alpha \beta}$ is diagonal and isotropic with $g^{\Vert}
=1.97$. Equivalent results can be obtained calculating the static
properties of model (\ref{Heisenberg}) via a transfer matrix
approach~\cite{TM_Pandit}. Since the uniaxial anisotropy $D$ is
rather strong with respect to the exchange coupling $\vert J
\vert$, as a first approximation one can assume the two sublattice
magnetizations to be directed just along the two easy axes,
$\hat{\mathbf{z}}_{2r-1}$ and $\hat{\mathbf{z}}_{2r}$, so that the chain
system (\ref{Heisenberg}) can be described by a non-collinear
Ising model formally identical to Eq. (\ref{non_collinear_H}),
with an effective~\cite{nota_angle} Ising exchange coupling $J_I$
and a generalized Land\'e factor $G_k$ defined as, respectively
\begin{eqnarray}
\label{bipartite} J_I&=&J S(S+1) \cos(\hat{\mathbf{z}}_{2r-1}
\cdot \hat{\mathbf{z}}_{2r})\cr
G_r&=&g_r^{\Vert}~\sqrt{S(S+1)}~(\hat{\mathbf{z}}_r \cdot
\hat{\mathbf{e}}_{\rm H}).
\end{eqnarray}

\begin{figure}\label{Mn}
\includegraphics[width=8cm,angle=0,bbllx=99pt,bblly=249pt,%
bburx=413pt,bbury=584pt,clip=true]{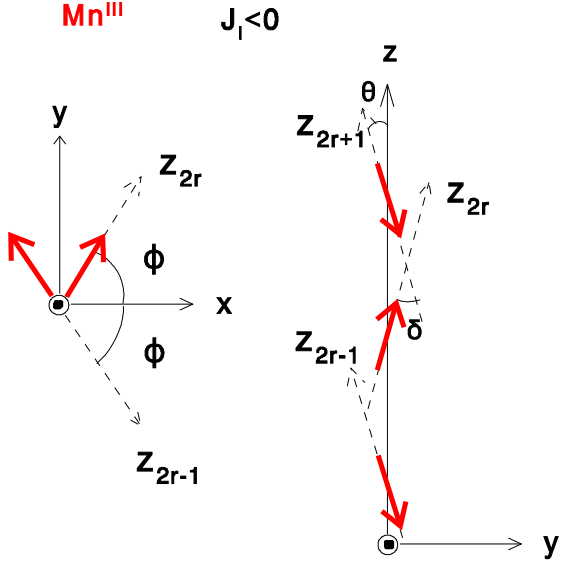} \caption{(color online)
Disposition of local axes ($\hat{\mathbf{z}}_{2r-1}$ and
$\hat{\mathbf{z}}_{2r}$) and magnetic moments (red arrows) in the
Mn$^{\rm III}$-based real SCM, discussed in Sect.
V.A, with antiferromagnetic effective Ising exchange coupling
$J_I<0$ . Right: Schematic view of the chain structure ($z$ is the
chain axis) along the crystallographic $x$ axis. Left: projections
of local axes (dashed lines) and of magnetic moments (red arrows)
in the $xy$ plane, perpendicular to the chain axis.}
\end{figure}

Depending on the orientation of the oscillating magnetic field
with respect to the crystallographic axes, the FT
of the generalized Land\'e factor takes the following forms
\begin{eqnarray}
\widetilde{G}_q&=&g^{\Vert}\sqrt{S(S+1)}~\cr &\times&{1\over
N}\sum_{r=1}^{N/2}e^{-i q 2r} [e^{iq}(\hat{\mathbf{z}}_{2r-1}
\cdot \hat{\mathbf{e}}_{\rm H}) +(\hat{\mathbf{z}}_{2r} \cdot
\hat{\mathbf{e}}_{\rm H})]  \cr &=& g^{\Vert} \sqrt{S(S+1)}
\begin{cases}
\sin \theta_1 \cos \phi_1 \delta_{q,0},~ {\rm H}\Vert x \\
\sin \theta_1 \sin \phi_1 \delta_{q,\pi},~ {\rm H}\Vert y  \\
\cos \theta_1 \delta_{q,0} ,~ {\rm H}\Vert z~ {\rm(chain~ axis)}
\end{cases}
\end{eqnarray}
The corresponding {\it a.c.} susceptibility takes the expression
\begin{eqnarray}
\chi(\omega, T) &=& N \mu^2_{B} \beta f(\beta J_I)(g^{\Vert})^2
[S(S+1)]\cr
  &\times&
\begin{cases}
\sin^2 \theta \cos^2 \phi {1\over {(1-\gamma_I)-i({{\omega}\over {\alpha}})}},~ {\rm H}\Vert x \\
\sin^2 \theta \sin^2 \phi {1\over {(1+\gamma_I)-i({{\omega}\over {\alpha}})}},~ {\rm H}\Vert y  \\
\cos^2 \theta {1\over {(1-\gamma_I)-i({{\omega}\over {\alpha}})}}
,~ {\rm H}\Vert z~ {\rm(chain~ axis)}
\end{cases}
\end{eqnarray}
Taking into account that, for the Mn$^{\rm III}$ SCM under study,
the ``true" exchange coupling, $J$ in Eq.~(\ref{Heisenberg}), is
antiferromagnetic, and that the angle between the two easy
anisotropy axes $\hat{\mathbf{z}}_1$ and $\hat{\mathbf{z}}_2$ is
$\delta=34.6^{\rm o}<90^{\rm o}$ (see Fig. 4, right), from Eq.
(\ref{bipartite}) it follows that also the effective Ising
exchange coupling is antiferromagnetic, $J_I<0$. As a consequence,
in the low temperature limit $\beta \vert J_I\vert \to \infty$,
the relaxation time $\tau_{q=\pi}$ diverges, while $\tau_{q=0}$
does not. Thus, for low frequencies $\omega/\alpha \ll 1$, the
{\it a.c.} susceptibility presents a resonant behavior only when
the oscillating magnetic field is applied along the
crystallographic $y$ axis, \textit{i.e.} the direction,
perpendicular to the chain axis, along which the magnetizations of
the two sublattices are {\it uncompensated} (see Fig. 4). In
contrast, when H is applied parallel to $z$ (the chain axis) or to
$x$, namely two directions along which the magnetizations of the
two sublattices are exactly compensated, no resonant behavior is
expected. These theoretical predictions turn out to be in
excellent agreement with experimental {\it a.c.} susceptibility
data~\cite{Montpellier} obtained in a single crystal sample of
[Mn(TPP)O$_2$PPhH]$\cdot$H$_2$O, thus confirming that such a
Mn$^{\rm III}$-based canted antiferromagnet is a {\it bona fide}
SCM.

\subsection{The Dy$^{\rm III}$-based Single Chain Magnet}
The molecular magnetic compound of formula
[Dy(hfac)$_3$(NITPhOPh)], hereafter denoted by Dy$^{\rm
III}$-based SCM, belongs to a family of quasi one-dimensional
magnets in which rare earth ions (with spin $S$) and organic
radical ions (with spin $s=1/2$) alternate themselves along the
chain axis, $z$, which in this compound coincides with the
crystallographic $b$ axis. Static measurements in single crystal
samples suggest~\cite{tesi_Kevin} that there is an
antiferromagnetic exchange interaction between neighboring
Dy$^{\rm III}$ ions, whose easy anisotropy axes are canted with
respect to the chain axis in such a way to generate an
uncompensated moment along $b$, while the components in the $ac$
plane are compensated.  Thus, as far as the dominant exchange
interaction $J<0$ between Dy$^{\rm III}$ ions is taken into
account, the spin Hamiltonian of the system is quite similar to
Eq.(\ref{Heisenberg}). However, with respect to the Mn$^{\rm
III}$-based chain, the crystal structure of the Dy$^{\rm
III}$-based SCM is more complicated, not only owing to the
presence of two kinds of magnetic centers (the Dy$^{\rm III}$ ions
and the organic radical ions), but mainly because the system is
formed by two different families of chains, with two almost
orthogonal projections of the easy axes in the $ac$ plane,
perpendicular to the chain axis: this ``accidental" (in the sense
that it is not imposed by symmetry) orthogonality is the reason
for the nearly isotropic magnetic behavior displayed by the system
within such a plane~\cite{tesi_Kevin}.

\begin{figure}\label{Dy}
\includegraphics[width=8cm,angle=0,bbllx=41pt,bblly=200pt,%
bburx=458pt,bbury=590pt,clip=true]{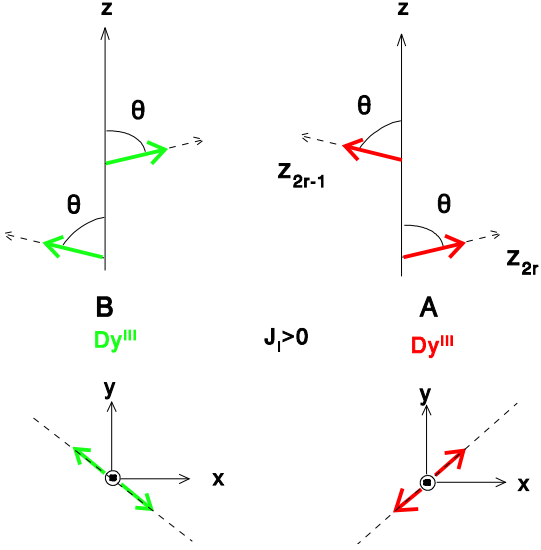} \caption{(color online)
Disposition of odd and even local axes ($\hat{\mathbf{z}}_{2r-1}$
and $\hat{\mathbf{z}}_{2r}$) and magnetic moments (thick arrows)
in the Dy$^{\rm III}$-based real SCM, discussed in
Sect. V.B, with ferromagnetic effective Ising exchange coupling
$J_I>0$ . Top: Schematic view of the chain structure ($z$ is the
chain axis), displaying the two families of chains (A, with red
magnetic moments, and B, with green magnetic moments). Bottom:
projections of magnetic moments in the $xy$ plane, perpendicular
to the chain axis.}
\end{figure}

We adopt a simplified model formally equivalent to Eq.
(\ref{Heisenberg}). Taking into account only the dominant
antiferromagnetic exchange interaction ($J<0$) between neighboring
Dy$^{\rm III}$ ions (which indeed are next nearest neighbors in
the real system) and their uniaxial anisotropy ($D>0$), the system
can approximately be described by the classical spins Hamiltonian
(\ref{Heisenberg}), where now $\vert S_k \vert=1$. By means of a
classical Transfer Matrix calculation, the static properties of
the Dy$^{\rm III}$-based SCM turn out to be satisfactorily
fitted~\cite{tesi_Kevin} by $J=-6$ K, $D=40$ K, $g^{\Vert}=10$,
with the two easy anisotropy axes $\hat{\mathbf{z}}_{2r-1}$,
$\hat{\mathbf{z}}_{2r}$ forming equal angles $\theta \approx
75^{\rm o}$ with the chain axis $z$. (Notice that the latter
property holds true for both families of chains.) Also in the case
of the Dy$^{\rm III}$-based SCM, the uniaxial anisotropy $D$ turns
out to be sufficiently strong with respect to the exchange
coupling $\vert J \vert$ in order to assume, as a first
approximation~\cite{nota_angle}, the two sublattice magnetizations
of Dy$^{\rm III}$ to be directed just along the two easy axes.
Thus one can define an equivalent non-collinear Ising model
(\ref{non_collinear_H}), where the effective Ising exchange
coupling $J_I$ and the generalized Land\'e factor $G_r$ are now
defined as
\begin{eqnarray}
\label{geiI}
J_I&=&J  \cos(\hat{\mathbf{z}}_{2r-1} \cdot
\hat{\mathbf{z}}_{2r})\cr G_r&=&g_r^{\Vert}~(\hat{\mathbf{z}}_r
\cdot \hat{\mathbf{e}}_{\rm H}).
\end{eqnarray}
Depending on the orientation of the oscillating magnetic field
with respect to the crystallographic axes, the FT
of the generalized Land\'e factor takes the form
\begin{eqnarray}
\widetilde{G}_q&=&g^{\Vert} ~{1\over N}\sum_{r=1}^{N/2}e^{-i q 2r}
[e^{iq}(\hat{\mathbf{z}}_{2r-1} \cdot \hat{\mathbf{e}}_{\rm H})
+(\hat{\mathbf{z}}_{2r} \cdot \hat{\mathbf{e}}_{\rm H})]
 \cr & \propto &  g^{\Vert}
 \begin{cases}
  \cos \theta ~\delta_{q,0},~~  {\rm H}\Vert z~~{\rm(chain~axis)} \\
 \sin \theta ~ \delta_{q,\pi},~~ {\rm H}\perp z
\end{cases}.
\end{eqnarray}
It is important to notice that this result holds true for both
families (A,B) of chains. Next, we observe that since in the
Dy$^{\rm III}$-based SCM, the spins on opposite sublattices are
coplanar with the chain axis, the angle between
$\hat{\mathbf{z}}_{2r-1}$ and $\hat{\mathbf{z}}_{2r}$ is just $2
\theta \approx 150^{\rm o}>90^{\rm o}$. Taking into account that
the ``true" exchange constant in Eq. (\ref{Heisenberg}) is
antiferromagnetic, $J<0$, from Eq. (\ref{geiI}) it follows that the effective Ising
exchange coupling is now ferromagnetic, $J_I>0$ (see Fig. 5, top). As a consequence, in the
low temperature limit $\beta J_I \to \infty$, the relaxation time
$\tau_{q=0}$ diverges, while $\tau_{q=\pi}$ does not. Thus, the
{\it a.c.} susceptibility
\begin{eqnarray}
\chi(\omega, T) &\propto& N \mu^2_{B} \beta f(\beta
J_I)(g^{\Vert})^2 \cr
  &\times&
\begin{cases}
\cos^2 \theta {1\over {(1-\gamma_I)-i({{\omega}\over {\alpha}})}},~ {\rm H}\Vert z  ~ {\rm(chain~ axis)}\\
\sin^2 \theta {1\over {(1+\gamma_I)-i({{\omega}\over {\alpha}})}}
,~ {\rm H}\perp z
\end{cases}
\end{eqnarray}
is expected to have a resonant behavior, for low frequencies
$\omega/\alpha \ll 1$, only when the oscillating magnetic field is
applied parallel to the chain axis, $z$, along which the
magnetizations of the two sublattices are {\it uncompensated} (see
Fig. 5, top). Such a theoretical prediction turns out to be in
excellent agreement with the experimental {\it a.c.}
susceptibility data\cite{tesi_Kevin} obtained in a single crystal
sample of [Dy(hfac)$_3$(NITPhOPh)]$_{\infty}$, thus confirming
that also the Dy$^{\rm III}$-based canted antiferromagnet is a
{\it bona fide} SCM. The only qualitative
difference, with respect to the Mn$^{\rm III}$-based chain is
that, due to the different geometry of the spin arrangement and of
the local anisotropy axes with respect to the chain axis, the
resonant behavior of the {\it a.c.} susceptibility is now observed
for field applied parallel to the chain axis, rather than
perpendicular to it.

\subsection{The CoPhOMe (Co$^{\rm II}$-based) Single Chain Magnet}
In the molecular magnetic compound of formula
[Co(hfac)$_2$NITPhOMe], hereafter denoted by
CoPhOMe~\cite{Caneschi_Angew,CoPhOMe_EPL}, the magnetic
contribution is given by Cobalt ions, with an Ising character and
effective $S=1/2$, and by NITPhOMe organic radical ions,
magnetically isotropic and with $s=1/2$.
The spins are
arranged on a helical structure, schematically depicted in Fig. 6, right, whose projections in a plane
perpendicular to the helix axis $z$ (coincident with the
crystallographic $c$ axis), are represented in Fig. 6, left. The
primitive magnetic cell is made up of three Cobalts (black arrows)
and three organic radicals (red arrows). Although the effective
spins of the two types of magnetic centers have the same value,
the gyromagnetic factors are different: $g_{Co} \ne g_R$; thus,
since the nearest neighbor (Cobalt-radical) exchange interaction
is negative (and strong, $\vert J \vert \approx 100$
K)~\cite{CoPhOMe_EPL}, the sublattice magnetizations are not
compensated along $z$, whereas they are compensated within the
$xy$ plane perpendicular to the chain axis $z$. For this compound,
which was the first to display SCM behavior~\cite{Caneschi_Angew,CoPhOMe_EPL},  
static measurements on single-crystal samples has not been interpreted
in terms of a simple model yet, due to the complexity of the system itself.
Thus, a relationship such as \eqref{bipartite} and \eqref{geiI}, which
associate the Ising Hamiltonian~\eqref{non_collinear_H} parameters ($J_I$ and $G_k$)
with those of a more realistic Hamiltonian, is still missing.
However, the dynamic behavior has been thoroughly investigated
treating - for the sake of simplicity -
both the Co$^{\rm II}$ and the
organic radical spins as Ising variables, with $\sigma=\pm1$.
The effective Ising Hamiltonian reads
\begin{eqnarray}
\label{Ising_Co} && \mathcal{H}=-\sum_{l=1}^{N/6}\sum_{m=1}^3 \,
\{
J_I \sigma_{l,2m}
\left[\sigma_{l,2m-1} + \sigma_{l,2m+1}  \right] + e^{-i\omega t} \mu_B {\rm H }
\cr && \left[ g_R \sigma_{l,2m-1} (\hat{\mathbf{z}}_{2m-1} \cdot
\hat{\mathbf{e}}_{\rm H}) +g_{Co} \sigma_{l,2m}
(\hat{\mathbf{z}}_{2m} \cdot \hat{\mathbf{e}}_{\rm H})\right] \}
\end{eqnarray}
with $l$ magnetic cell index and $m$ site label with boundary conditions $\sigma_{l,7}=\sigma_{l+1,1}$. 
\begin{figure}\label{Cophome}
\includegraphics[width=8cm,angle=0,bbllx=88pt,bblly=256pt,%
bburx=555pt,bbury=619pt,clip=true]{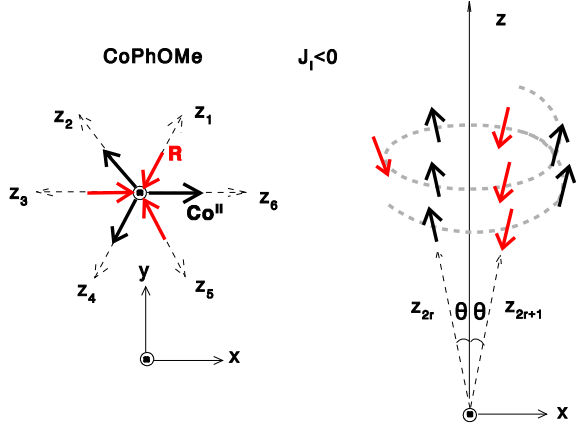} \caption{(color
online) Disposition of even and odd local axes (dashed lines) and
magnetic moments (thick arrows) in the CoPhOMe real SCM,
discussed in Sect. V.C, with antiferromagnetic effective
Ising exchange coupling $J_I<0$ . Right: Schematic view of the
chain structure ($z$ is the chain axis) along the crystallographic
$y$ axis. Left: projections of local axes (dashed lines) and of
magnetic moments (thick arrows) in the $xy$ plane, perpendicular
to the chain axis.}
\end{figure} 
Since all the local axes $\hat{\mathbf{z}}_{k}$ ($k=1,\cdots,6$)
form the same angle $\theta \approx 55^{\rm o}$ with the $z$ axis,
when a magnetic field is applied along $z$, the FT
of the generalized Land\'e factor is simply given by
\begin{eqnarray}
\label{generalized} &&\widetilde{G}_q^{\Vert}=\cos\theta{1\over
N}\sum_{r=1}^{N/2} (g_{Co}e^{-iq(2r-1)}+g_R e^{-iq 2r})\cr
&&={{\cos\theta}\over 2}
[(g_{Co}+g_R)\delta_{q,0}+(g_{Co}-g_R)\delta_{q,\pi}]
\end{eqnarray}
\noindent which, except for the prefactor $\cos\theta$, is quite
similar to Eq.~(\ref{collinear_altern}) for the collinear Ising
chain with alternating spins. Thus, the parallel \textit{a.c.}
susceptibility is
\begin{eqnarray} \label{ac_Susceptibility_par_CoPhOMe_Glauber}
\chi_{\Vert}(\omega, T) &=& N \mu^2_{B}\beta~  f(\beta
J_I){{\cos^2\theta}\over 4}\cr &\times& [\frac{
(g_{Co}+g_R)^2}{(1-\gamma_I) -i ({{\omega}\over {\alpha}})}
+\frac{(g_{Co}-g_R)^2}{(1+\gamma_I) -i ({{\omega}\over {\alpha}})}
]\,.
\end{eqnarray}
Taking into account that the effective exchange coupling of
CoPhOMe is negative ($J_I<0$), the antiferromagnetic branch of the
parallel susceptibility is characterized by a diverging time scale
$\tau_{q=\pi}=[\alpha(1+\gamma_I)]^{-1}$ at low temperature, so
that, for low frequencies $\omega/\alpha \ll 1$,
$\chi_{\Vert}(\omega, T)$ displays a resonant behavior.

Let us now consider the case of a field applied in the plane
perpendicular to the chain axis. For H$\Vert x$ (see Fig. 6, left) one
has, letting $k_0={{\pi}\over 3}$
\begin{eqnarray}
&&G^x_{2r-1}=g_R \sin\theta \cos\left[k_0(2r-1) \right]\cr
&&G^x_{2r}=g_{Co} \sin\theta \cos\left[k_0 2r \right]
\end{eqnarray}
so that the FT takes the form
\begin{eqnarray}
\widetilde{G}^x_q&=& \sin\theta {1\over {N}}\sum_{r=1}^{N/2}
e^{-iq 2r}\Big( g_{Co}\cos(k_0 2r) \cr &+&g_R e^{iq} [\cos(k_0)
\cos(k_0 2r) + \sin(k_0) \sin(k_0 2r)]\Big)\cr &=&{1\over
4}\sin\theta [(g_{Co}+g_R
e^{i(q-k_0)})(\delta_{q,k_0}+\delta_{q,\pi+k_0})\cr &+&
(g_{Co}+g_R e^{i(q+k_0)})(\delta_{q,-k_0}+\delta_{q,\pi-k_0})]
\end{eqnarray}
where, as usual, we have exploited Eq.~(\ref{boundary}). Thus it
follows that
\begin{equation}
\begin{split}
&\widetilde{G}_{q=\pm \frac{\pi}{3}}=\frac{\sin\theta}{4}\left( g_{Co} + g_R \right)\\
&\widetilde{G}_{q=\pi \pm
\frac{\pi}{3}}=\frac{\sin\theta}{4}\left( g_{Co} - g_R \right).
\end{split}
\end{equation}
The corresponding relaxation times are $\tau_{q=\pm
\frac{\pi}{3}}=\frac{\alpha}{1-\frac{1}{2} \gamma}$ and $\tau_{q=\pm
\frac{2\pi}{3}}=\frac{\alpha}{1+\frac{1}{2} \gamma}$ so that,
summing the four contributions we obtain the perpendicular
{\it a.c.} susceptibility
\begin{eqnarray}
 \label{ac_Susceptibility_perp_CoPhOMe_Glauber}
 &&\chi_{\perp}(\omega, T) =N \mu^2_{B}\beta~f(\beta J_I) \frac{ \sin^2\theta }{8} \cr
 &\times& \left[
\frac{(g_{Co}+g_R)^2}{(1-\frac{1}{2} \gamma)
 -i (\frac{\omega}{\alpha})}  +
\frac{(g_{Co}-g_R)^2}{(1+\frac{1}{2} \gamma)
 -i (\frac{\omega}{\alpha})} \right].
\end{eqnarray}

In conclusion, for the six-fold helix model with alternating spins
and Ising exchange coupling in Eq.~(\ref{Ising_Co}), the parallel
and perpendicular components of the {\it a.c.} susceptibility,
$\chi_{\Vert}(\omega, T)$ and 
$\chi_{\perp}(\omega, T)$, display a behavior similar to that
of a ferrimagnetic chain with alternating spins
 (see \eqref{ac_Susceptibility_Ferrimagnet_Glauber}) and
of an $n$-fold helical spin arrangement with equivalent spins
 (see \eqref{ac_Susceptibility_perp_helix_Glauber}), respectively.
In spite of the approximations involved in model (\ref{Ising_Co})
to describe the real CoPhOMe molecular magnetic chain, the two
calculated susceptibilities
\eqref{ac_Susceptibility_par_CoPhOMe_Glauber} and
\eqref{ac_Susceptibility_perp_CoPhOMe_Glauber}, qualitatively
reproduce the dynamic behavior of this compound~\cite{Caneschi_Angew,CoPhOMe_EPL}.
In fact, no out-of-phase
\textit{a.c.} susceptibility (imaginary part) is observed when the
field is applied in the plane perpendicular to the chain axis,
$z$, for the experimental frequencies ($1 \div 10^5$ Hz)~\cite{CoPhOMe_EPL}.
In contrast, when the oscillating field is
parallel to $z$, a resonant behavior is observed as a function of
temperature. Even though our theoretical treatment holds only for
small deviations from equilibrium, it is worth mentioning that the
absence of slow relaxation for fields applied in the perpendicular
plane is evidenced in the low temperature magnetization curve as
well: at low enough temperatures, a finite-area hysteresis loop is
present only when a static field is applied parallel to the chain
axis, while no hysteresis is observed in the in-plane
magnetization curve~\cite{Caneschi_Angew,CoPhOMe_EPL}.

\section{Conclusions}
In conclusion, in the framework of a one-dimensional
Ising model with single spin-flip Glauber dynamics,
taking into account reciprocal non-collinearity of local anisotropy axes
and the crystallographic (laboratory) frame, we have investigated: (i)
the dynamics of magnetization reversal in zero field, and (ii) the response
of the system to a weak magnetic field, oscillating in time.
We have shown that SCM behavior is not only a feature of
collinear ferro- and ferrimagnetic, but also of canted
antiferromagnetic chains. In particular, we have found that
resonant behavior of the {\it a.c.} susceptibility versus temperature in
response to an oscillating magnetic field is possible, at low frequency,
only for fields applied in a direction where magnetic moments are uncompensated.
In contrast, a broad peak is expected when there is no net magnetization
along the field direction.

The role played by geometry in selecting the time scales involved
in a process is an important and peculiar result,
typical of magneto-molecular approach to low-dimensional
magnetism. 
In fact, magnetic centers with uniaxial anisotropy
usually correspond to  building blocks with low symmetry~\cite{Gatteschi_Science,Gatteschi_Sessoli_Review}, 
which -- in turn -- often crystallize in more symmetric
space groups, realizing a reciprocal non-collinearity between local anisotropy axes
as a natural consequence~\cite{Review_CCM,Miyasaka}.
Thus the family of real SCM's, to which
our model applies, does not restrict to
\textit{ad-hoc} synthesized compounds but, instead,
is expected to grow larger in the future~\cite{feature_SCM}.
As a validity check of our selection rules (as well as a tutorial
exemplification), we have shown how our theory applies successfully to three different
molecular-based spin chains; when possible, we have put the
parameters of our model Hamiltonian~\eqref{non_collinear_H} in relationship with
those of more general models, typically used to fit the static
properties of the corresponding compounds. 
Needless to say that the possibility of schematizing the chosen three 
compounds with Hamiltonian~\eqref{non_collinear_H} relies on the 
fact that at low enough temperatures they behave as chains consisting of two-level units 
coupled by a fully anisotropic exchange interaction. 
The latter assumption is expected to hold 
also for spin larger than one-half 
in the presence of strong single ion anisotropy, provided 
that domain walls still remain sharp~\cite{Ale_per_Gatto_2008,nota_angle}. 
In this case each single magnetic center follows a thermally activated 
dynamics, with an energy barrier $\Delta_0$, and well established 
heuristic arguments~\cite{Clerac_Coulon} suggest to replace the attempt frequency 
by  $\alpha=\alpha_0 e^{-\beta\Delta_0}$. \\
A naive application of our 3-fold-helix 
results~\eqref{ac_Susceptibility_zz_helix2_Glauber} 
and~\eqref{ac_Susceptibility_perp_helix_Glauber} 
to the recently synthesized non-collinear Dy$_{3}$ cluster would prevent the observation of 
\textit{slow} relaxation, while Single-Molecule-Magnet dynamics is indeed there observed 
even in the presence of compensated magnetic moments~\cite{Javier_PRL}. 
However, such a behavior in the classical regime, \textit{i.e.}   
far from level crossings where 
underbarrier processes of quantum origin are important, is 
observed for Dy$_3$ in non-zero field and 
the resonant behavior is due to a change of the relative population  
between the lowest and the first excited Kramers doublets of 
each Dy ion: For sure this mechanism cannot be accounted for when  
dealing with two-level elementary variables, like $\sigma_k$ in  Hamiltonian~\eqref{non_collinear_H}. 
An extension of our model to multivalued 
$\sigma_k$'s definitely deserve to be considered in the 
next future. 

Beyond molecular spin chains, our approach
might also be used to model monatomic nanowires showing slow
relaxation of the magnetization at low
temperatures~\cite{Gambardella} and, possibly, one-dimensional
spin glasses~\cite{Mydosh} (provided that quenched disorder is
somehow taken into account). In this regard, the question of
distinguishing between SCM and spin-glass behavior in quasi-1D
systems is still a hot topic of
discussion~\cite{Maignan,Etzkorn,Bogani_ICMM,Girtu}.

After the successful organization of
Single-Molecule Magnets onto surfaces~\cite{A_Cornia,Abdi,Mannini}, 
the grafting of properly functionalized SCM's on substrates 
represents a foreseeable goal as well as a fundamental step for their possible use
as magnetic-memory units~\cite{feature_SCM}.
Technologies employing more traditional materials but based on alternative geometrical arrangement of magnetic
anisotropy axes with respect to the switching field,
such as in perpendicular recording~\cite{Piramanayagam} or processional switching~\cite{C_Back},
are already at the stage of forthcoming implementation in devices.
Were SCM's to be considered as a possible route to tackle the main issues of high-density magnetic storage
-- \textit{i.e.}  optimization of the signal-noise ratio, thermal stability and writability~\cite{Piramanayagam} --
the proposed selection rules for slow relaxation, and related bistability,
might find an application in magnetic-memory manufacture as well.

\acknowledgments We wish to thank R. Sessoli and A. Rettori
for stimulating discussions, and J. Villain for interest and
fruitful suggestions in the early stages of this research work.
Financial support from ETHZ, the Swiss National Foundation, and
Italian National Research Council is also acknowledged.

\appendix
\section{The Generating Functions approach}
Assuming periodic boundary conditions and defining the two
generating functions
\begin{eqnarray}
\label{LGdefinition}  \mathcal{L} (y ,t)&=&\sum ^{+\infty
}_{r=-\infty }y ^{2r+1}\, s_{2r+1}(t) \cr \mathcal{G} (y
,t)&=&\overset{+\infty }{\underset{r=-\infty}\sum} y^{2r}\,
s_{2r}(t)\,,
\end{eqnarray}
Eqs.~(\ref{newform}) take the form of two differential equations
(with the dot indicating the first derivative with respect to the
adimensional variable $\alpha t$)
\begin{equation}
\label{Genrating_system}
\begin{cases}
\dot{\mathcal{L}} (y ,t)=-\mathcal{L} (y ,t)+\frac{1}{2}\gamma (y
+y ^{-1}) \mathcal{G} (y ,t)
\\
\dot{ \mathcal{G}}  (y ,t)=- \mathcal{G} (y ,t)+\frac{1}{2}\gamma
(y +y^{-1})\mathcal{L} (y ,t)\,.
\end{cases}
\end{equation}

The system \eqref{Genrating_system} can be decoupled through the
substitution
\begin{eqnarray}
\label{substitution} \mathcal{U}(y ,t)&=&\mathcal{L}
(y,t)+\mathcal{G} (y ,t)\cr \mathcal{W}(y ,t)&=&\mathcal{L} (y
,t)-\mathcal{G} (y ,t)\,,
\end{eqnarray}
from which we directly get
\[
\begin{cases}
\dot{\mathcal{U}}(y ,t)=-(1-\nu )\mathcal{U}(y ,t)\\
\dot{\mathcal{W}}(y ,t)=-(1+ \nu )\mathcal{W}(y ,t)\,,
\end{cases}
\]
with $\nu =\frac{1}{2}\gamma (y +y ^{-1}) $. The solutions of
these two equations are $\mathcal{U}(y ,t)=\mathcal{U}(y
,0)e^{-(1-\nu )\alpha t}$ and $\mathcal{W}(y ,t)=\mathcal{W}(y
,0)e^{-(1+\nu )\alpha t}$ that, exploiting the property
\begin{equation}
\label{Bessel_Prop} \exp \left[ \frac{1}{2}(y +y ^{-1})x\right]
=\sum ^{+\infty }_{k=-\infty }y ^{k}\, \mathcal{I}_{k}(x)
\end{equation}
of the Bessel functions of imaginary argument $ \mathcal{I}_{k}(x)
$, can be rewritten as
\[
\begin{cases}
\mathcal{U}(y ,t)=\mathcal{U}(y ,0)e^{-\alpha t} \overset{+\infty
}{\underset{k=-\infty}\sum} y ^{k}\mathcal{I}_{k}(\gamma \alpha
t)\\ \mathcal{W}(y ,t)=\mathcal{W}(y ,0)e^{-\alpha
t}\overset{+\infty }{\underset{k=-\infty}\sum} (-1)^{k}y
^{k}\mathcal{I}_{k}(\gamma \alpha t)
\end{cases}
\]
Performing the inverse transformation of \eqref{substitution}, the
solutions to the system \eqref{Genrating_system} can be obtained
\begin{eqnarray}
\nonumber &&\mathcal{L}(y ,t)=\frac{1}{2}e^{-\alpha t}\times \cr
&&\overset{+\infty }{\underset{k=-\infty}\sum} y ^{k}\, \left[
\mathcal{U}(y ,0)\mathcal{I}_{k}(\gamma \alpha
t)+(-1)^{k}\mathcal{W} (y ,0)\mathcal{I}_{k}(\gamma \alpha
t)\right] \cr\cr &&\mathcal{G}(y ,t)=\frac{1}{2}e^{-\alpha
t}\times \cr &&\overset{+\infty }{\underset{k=-\infty}\sum} y
^{k}\, \left[ \mathcal{U}(y ,0)\mathcal{I}_{k}(\gamma \alpha
t)-(-1)^{k}\mathcal{W}(y ,0)\mathcal{I}_{k}(\gamma \alpha
t)\right]\,.
\end{eqnarray}
Then, separating the $k$-odd from the $k$-even  terms in both
sums, we get
\begin{eqnarray}
\nonumber &&\mathcal{L}(y ,t)=e^{-\alpha t}\overset{+\infty
}{\underset{r=-\infty}\sum} [y ^{2r}\mathcal{L}(y
,0)\mathcal{I}_{2r}(\gamma \alpha t)\cr &&+y ^{2r+1} \mathcal{G}(y
,0)\mathcal{I}_{2r+1}(\gamma \alpha t)]\cr
 &&\mathcal{G}(y ,t)=e^{-\alpha t}\overset{+\infty }{\underset{r=-\infty}\sum}[ y
^{2r}\mathcal{G}(y ,0)\mathcal{I}_{2r}(\gamma \alpha t)\cr &&+y
^{2r-1}\mathcal{L}(y ,0)\mathcal{I}_{2r-1}(\gamma \alpha t)]\,.
\end{eqnarray}

According to \eqref{LGdefinition}, now $\mathcal{L}(y ,0) $ and
$\mathcal{G}(y ,0) $ can be expressed again in terms of the
initial single spin expectation values $s_{2r}(0)$ and
$s_{2r+1}(0)$ respectively
\begin{eqnarray}
\nonumber &&\mathcal{L}(y ,t)=e^{-\alpha t} \sum ^{+\infty
}_{r=-\infty }\cr &&\times [ y ^{2r}  \sum ^{+\infty }_{m=-\infty
}y ^{2m+1}s_{2m+1}(0) \mathcal{I}_{2r}(\gamma \alpha t)\cr && +y
^{2r+1} \sum ^{+\infty }_{m=-\infty }y ^{2m}s_{2m}(0)
\mathcal{I}_{2r+1}(\gamma \alpha t) ]
\end{eqnarray}
Putting $k'=k+m$ we have
\begin{eqnarray}
\nonumber &&\mathcal{L}(y ,t)=e^{-\alpha t} \sum ^{+\infty
}_{k'=-\infty }y ^{2k'+1} \sum ^{+\infty }_{m=-\infty }\cr && [
s_{2m+1}(0)\mathcal{I}_{2(k'-m)}(\gamma \alpha t)
+s_{2m}(0)\mathcal{I}_{2(k'-m)+1}(\gamma \alpha t)]\,.
\end{eqnarray}
Comparing  this latter result with the definition of
$\mathcal{L}(y ,t)$ \eqref{LGdefinition} and requiring for the
terms corresponding to the same power of $y$ to be equal, an
explicit function for the odd spin expectation values is readily
obtained
\begin{eqnarray}
\label{odd_sites_relax_Solution}&& s_{2r+1}(t)=e^{-\alpha t} \sum
^{+\infty }_{m=-\infty }[ s_{2m+1}(0)\mathcal{I}_{2(r-m)}(\gamma
\alpha t)\cr &&+s_{2m}(0)\mathcal{I}_{2(r-m)+1}(\gamma \alpha
t)]\,.
\end{eqnarray}
Substituting 
$\mathcal{L}(y,0) $ e $\mathcal{G}(y ,0) $  in the solution found for
$\mathcal{G}(y,t) $ and performing similar passages, we obtain
the expectation value for even sites
\begin{eqnarray}
\label{even_sites_relax_Solution}&& s_{2r}(t)=e^{-\alpha t}\sum
^{+\infty }_{m=-\infty }[ s_{2m}(0)\mathcal{I}_{2(r-m)}(\gamma
\alpha t)\cr&&+s_{2m+1}(0)\mathcal{I}_{2(r-m)-1}(\gamma \alpha
t)]\,.
\end{eqnarray}


\begin{thebibliography}{199}
\bibitem{nota_equil}
It is worth noticing that, once a good candidate were found, an
experimental verification of the theoretical prediction might be
not so easy since the considered relaxation is a strongly
out-of-equilibrium process, while in our theory small departures
from equilibrium were assumed.
\end{thebibliography}
\end{document}